# Near-Unity Emitting, Widely Tailorable and Stable Exciton Concentrators Built from Doubly Gradient 2D Semiconductor Nanoplatelets

Xiao Liang,[1] Emek G. Durmusoglu,[1,2] Maria Lunina,[3] Pedro Ludwig Hernandez-Martinez,[1] Vytautas Valuckas,[4] Fei Yan,[1] Yulia Lekina,[2] Vijay Kumar Sharma,[1] Tingting Yin,[2] Son Tung Ha,[4] Ze Xiang Shen,[2] Handong Sun,[2] Arseniy Kuznetsov,[4] and Hilmi Volkan Demir[1,2,5]\*

[1]LUMINOUS! Center of Excellence for Semiconductor Lighting and Displays, The Photonics Institute, School of Electrical and Electronic Engineering, Nanyang Technological University, Singapore, 639798, Singapore

[2]Division of Physics and Applied Physics, School of Physical and Mathematical Sciences, Nanyang Technological University, Singapore, 637371, Singapore

[3]Interdisciplinary Graduate Program, Nanyang Technological University, Singapore, 637371 Singapore

[4]Institute of Materials Research and Engineering, A*STAR (Agency for Science, Technology and Research), 2 Fusionopolis Way, #08-03 Innovis, 138634, Singapore

[5]UNAM—Institute of Materials Science and Nanotechnology, The National Nanotechnology Research Center, Department of Electrical and Electronics Engineering, Department of Physics, Bilkent University, Bilkent, Ankara, 06800, Turkey

KEYWORDS: semiconductor nanoplatelets, near-unity quantum yield, tailorable electrostatic interactions, high stability, optoelectronics

ABSTRACT: The strength of electrostatic interactions (EI) between electrons and holes within semiconductor nanocrystals profoundly impact the performance of their optoelectronic systems, and different optoelectronic devices demand distinct EI strength of the active medium. However, achieving a broad range, fine-tuning of the EI strength for specific optoelectronic applications is a daunting challenge, especially in quasi 2-dimensional core-shell semiconductor nanoplatelets (NPLs), as the epitaxial growth of the inorganic shell along the direction of the thickness that solely contributes to the quantum confined effect significantly undermines the strength of the EI. Herein we propose and demonstrate a novel doubly-gradient (DG) core-shell architecture of semiconductor NPLs for on-demand tailoring of the EI strength by controlling the localized exciton concentration *via* in-plane architectural modulation, demonstrated by a wide tuning of radiative recombination rate and exciton binding energy. Moreover, these exciton-concentration-engineered DG NPLs also exhibit a near-unity quantum yield, remarkable thermal and photo stability, as well as considerably suppressed self-absorption. As proof-of-concept demonstrations, highly efficient color converters and high-performance light-emitting diodes (external quantum efficiency: 16.9%, maximum luminance: 43,000 cd/m$^2$) have been achieved based on the DG NPLs. This work thus opens up new avenues for developing high-performance colloidal optoelectronic device applications.

INTRODUCTION

Optoelectronic materials, possessing unique strength and dynamics of electrostatic interactions between electrons and holes,[1-3] have unlocked the potential for a wide range of cutting-edge optoelectronic applications, such as light-emitting diodes (LEDs),[4-6] lasers,[7-9] and quantum information technologies.[10, 11] Within the diverse families of optoelectronic materials, quasi 2-dimensional (2D) II-VI semiconductor nanoplatelets (NPLs), also known as colloidal quantum wells, comprised of cadmium chalcogenides (CdX, where X can be Se, S, or a combination thereof) within a zinc blende structure, have attracted increasing attention thanks to their ability to precisely control thickness at the atomic level (typically 3, 4 or 5 monolayers),[12] enabling uniform and strong 1D quantum confinement effects with remarkably large absorption cross-sections, high exciton binding energies ($E_b$), and narrow emission linewidths.[13-15] Despite the exceptional properties of atomically flat 2D NPLs, their practical applications have been hindered by their poor stability of quantum efficiency and low tolerance for surface defects, attributed to the unstable nature of the organic ligands used for surface passivation.[16] To address this challenge, synthetic strategies have been developed, which utilize stable inorganic semiconductors with a wider energy bandgap as surface passivation layer for encapsulating core or core-crown NPLs with shells, enabling a significant improvement in both stability and quantum efficiency.[17, 18] However, the epitaxial growth of the inorganic shell along the direction of the thickness that solely contributes to the quantum confined effect unavoidably undermines the strength of the electrostatic interactions. The aim of independently and selectively tailoring $E_b$, the energy required to bind an electron and a hole through the electrostatic Coulomb force, while simultaneously maintaining near-unity quantum yield and outstanding stability in existing core-shell (C@S) structures, has been a formidable challenge thus far.[19-21]

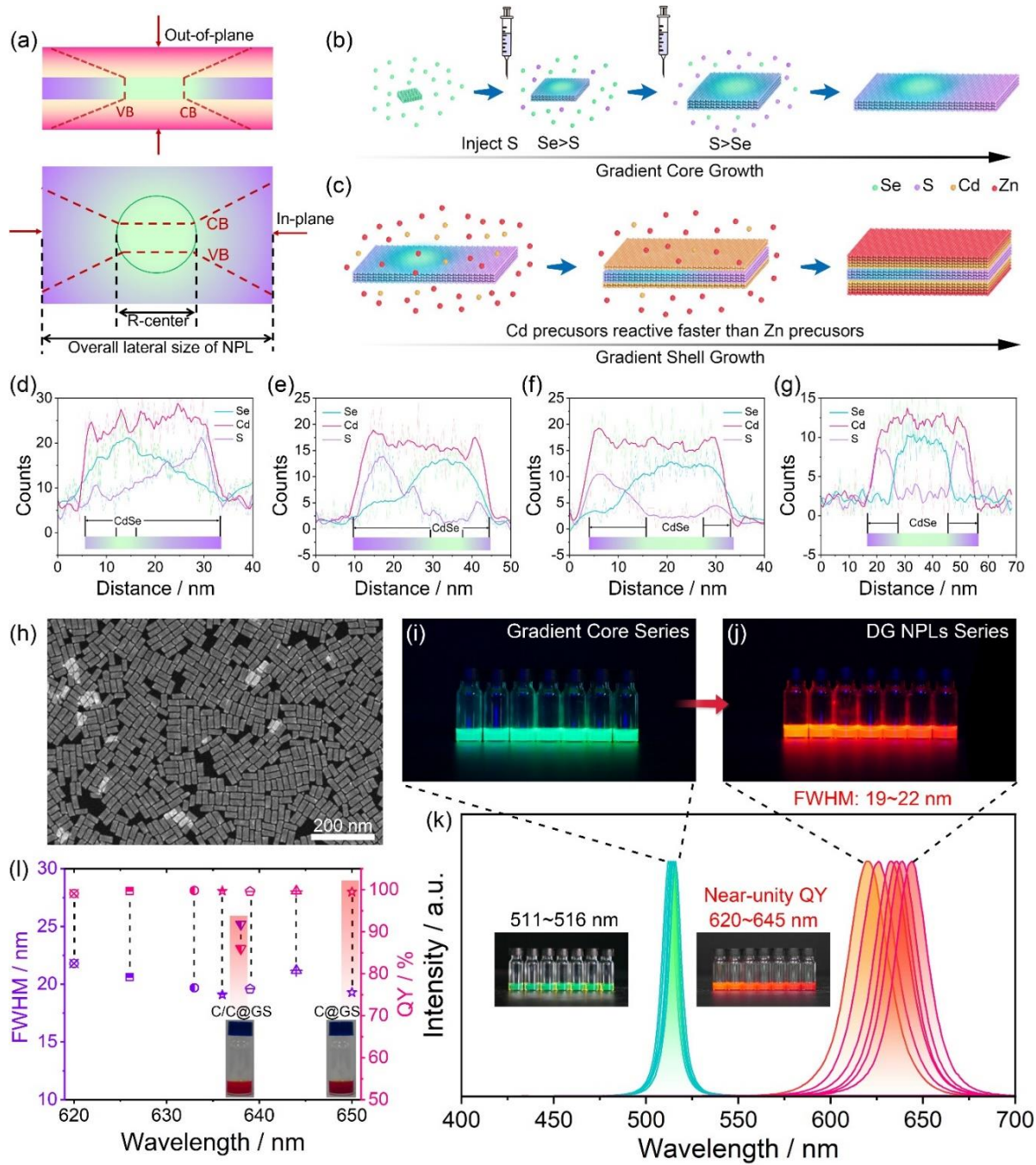

**Figure 1.** Architectural design, synthetic strategy, and characterizations of DG NPLs. (a) Architectural concept for the proposed doubly gradient NPLs using composition and potential barrier gradients in both out-of-plane (top) and in-plane (bottom) directions. CB and VB refer to conductive and valence band, respectively. Synthetic methods for the (b) one-pot construction of gradient core and (c) hot-injection method of gradient shell growth. EDS line-scan results for gradient cores with (d) small, (e) medium, (f) large size of CdSe seeds and (g) conventional CdSe/CdS core/crown NPLs. (h) Representative HAADF-STEM image of the DG NPLs with small a size of CdSe seed. (i-j) Photographs and (k) emission spectra for the gradient cores and DG NPLs series. (l) FWHM and quantum yields of DG NPLs series in comparison to conventional C@GS and C/C@GS NPLs.

The fine-tuning of electrostatic Coulomb force between electrons and holes, known as $E_b$, is paramount in optoelectronic systems with varying functionalities, as they demand distinct EI strength of the active medium.[1,3] For instance, judiciously tuning the $E_b$ in LEDs facilitates exciton generation and recombination according to a sequential electron-hole injection mechanism and boosts electroluminescence (EL) performance.[22,23] Moreover, strengthened electrostatic interactions enable faster radiative recombination rates in optoelectronic devices, leading to faster switching times and higher data rates in light-based communication.[24,25] Furthermore, there exists a strong desire for an increased $E_b$ in the realm of exciton-polariton lasers, as this facilitates robust polariton condensation even at higher temperatures.[26-28] Therefore, to unlock the full potentials of NPLs and facilitate their integration into advanced optoelectronic devices, it is critically important to tackle the challenge of reconciling the conservation of their exceptional stability achieved through thick shells with the fine-tuning of their exciton binding energy.

In addition to the target of on-demand tuning the strength of electrostatic interactions, state-of-the-art C@S NPLs also suffer from significant self-absorption,[29,30] resulting from the high spatial overlap of exciton generation and recombination, which further limits their efficiency in practical applications. While core-crown-shell (C/C@S) NPLs have partially mitigated this issue by delocalizing electrons to the crown while retaining holes in the core, their reported quantum yield typically falls short of 87% due to non-radiative trap states at the core-crown interface.[31-33]

To address the limitations of state-of-the-art NPLs mentioned above, herein we propose a novel concept of doubly-gradient (DG) architecture with a gradient core-gradient shell configuration, serving as a tailorable exciton concentrator, as illustrated in **Figure 1a**. Essentially, the proposed architecture operates as a highly efficient light harvester, leveraging on the large absorption coefficient of NPLs, and the designed type-I band offsets from both the in-plane of the gradient core and the out-of-plane of the gradient shell enable high concentrations of the generated excitons towards the recombination center through carrier transfers. Therefore, a large modulation of electrostatic interactions between electrons and holes can be achieved by engineering the localized exciton concentration *via* tuning the relative lateral dimension of the recombination center (R-center) compared to the overall NPL without altering the shell thickness that is required for high quantum efficiency and high stability. Moreover, this carrier transfer process also spatially separates the exciton generation and recombination, which is beneficial for mitigating the severe self-absorption issue observed in current core-shell NPLs.

To experimentally realize the proposed DG architecture, we construct the gradient core from a CdSe seed with an in-plane compositional gradient transition towards CdS (CdSe/CdSe$_x$S$_{1-x}$), while the gradient shell consists of an out-of-plane compositional gradient transition from CdS to ZnS (Cd$_x$Zn$_{1-x}$S), enabling bidirectional type-I band alignments. Our systematic photophysical studies demonstrate that these DG NPLs function as highly efficient exciton concentrators with super-fast and efficient carrier transfer dynamics facilitated by type-I band alignment and smooth interface transitions with fewer defects, leading to a near-unity quantum yield that is unachievable from conventional C/C@S structures. Remarkably, these exciton-concentration-engineered DG NPLs with different CdSe seed sizes demonstrate tunable emission wavelengths ranging from 620 to 645 nm due to the different in-plane spatial confinement, while maintaining near-unity quantum yields and narrow full-width at half-maximum (FWHM) in the range of 19-22 nm. Additionally, significantly reduced self-absorption is achieved in these DG NPLs by up to 59.2% compared to core-gradient shell (C@GS) counterparts. Furthermore, the $E_b$ and radiative recombination rates are enhanced by up to 103.2% and 54.9%, respectively, compared to C@GS NPLs. Moreover, the higher $E_b$ as well as the effective passivation of surface defects together contribute to the extraordinary stability of DG NPLs against thermal and UV degradation. Finally, micro-sized fluorescent patterns composed of these highly efficient and stable DG NPLs, as well as LED devices based on the $E_b$-tailored DG NPLs have been shown as proof-of-concept demonstrations. These DG NPLs-based LED devices exhibit an improved external quantum efficiency (EQE) of 16.9% and a maximum luminance of 43,000 cd/m$^2$ compared to C@GS NPLs with the same aspect ratio (EQE: 11.2%, maximum luminance: 14,100 cd/m$^2$). In summary, these novel DG NPLs with a near-unity quantum yield, tunable emission wavelength, narrow FWHM, significantly reduced self-absorption, tailorable $E_b$, and remarkable thermal and photo stability hold great promise for various practical applications of optoelectronics (color conversion, LEDs, polariton lasers, and so on).

RESULTS AND DISCUSSION

The proposed doubly gradient-structured NPLs were developed from a stepwise synthetic strategy as depicted in **Figure 1b-c.** The growth of gradient core NPLs originated from the lateral expansion of 4-monolayer (4ML) CdSe seed NPLs, unlike the synthesis of previously reported core-crown NPLs, where the CdSe core and CdS crown were grown separately,[34, 35] the gradient transition from CdSe to CdS was achieved by introducing controlled amounts of sulfur (S) precursors into the same pot, at a specific rate, after the CdSe seeds had grown to a certain size. As the Se precursor was continually depleted and S was added, the gradient core gradually transformed from CdSe to CdS (**Figure 1b**). This one-pot synthetic approach enables precise tuning of CdSe seed dimensions over a broad range and facile in-plane architectural engineering of the resultant gradient core heterostructure. Subsequently, to create the gradient shell, the dissimilar reaction rates of Zn and Cd precursors with the S precursors were leveraged,[36, 37] as presented in **Figure 1c**. The quicker reaction rate between Cd precursors and S generated a primary CdS shell, while the slower participation of Zn precursors occurred at later stages of the reaction, gradually forming the outermost layer of ZnS.[37] Transmission electron microscopy (TEM) images (**Figure S1a-c**) reveal the quasi-rectangular shapes of the gradient cores with distinct initiation injection times of S precursors of 100, 200, and 300 s, respectively. The corresponding energy-dispersive X-ray spectroscopy (EDS) line-scan results in **Figure 1d-f** reveal an increased size of CdSe seed from about 4 to 8 nm and then 11 nm along the long axis of the NPL, denoted as small-, medium-, and large-gradient cores, respectively. Additionally, the EDS line-scan results also demonstrate the compositional gradient and smooth transition from CdSe to CdS, as expected, in all gradient cores. In contrast, the conventional CdSe/CdS core-crown NPLs exhibit a larger CdSe size of approximately 15 nm, with a sharp interface between CdSe and CdS (**Figure 1g**), consistent with the literature.[34, 38] Following the shell growth, the resulting DG NPLs series, indicated as small-, medium-, and large-DG NPLs, built from their corresponding gradient cores, exhibit highly homogenous coating of $Cd_xZn_{1-x}S$ shells as well as uniform rectangular morphologies with similar lateral dimensions of length (35 nm) and width (15 nm) (size distribution results are summarized in **Figure S3)** and an average aspect ratio of around 2.3, under the optimized experimental conditions, as shown from the high-angle annular dark-field scanning TEM (HAADF-STEM) images in **Figure 1h** and TEM images in **Figure S2a-c**.

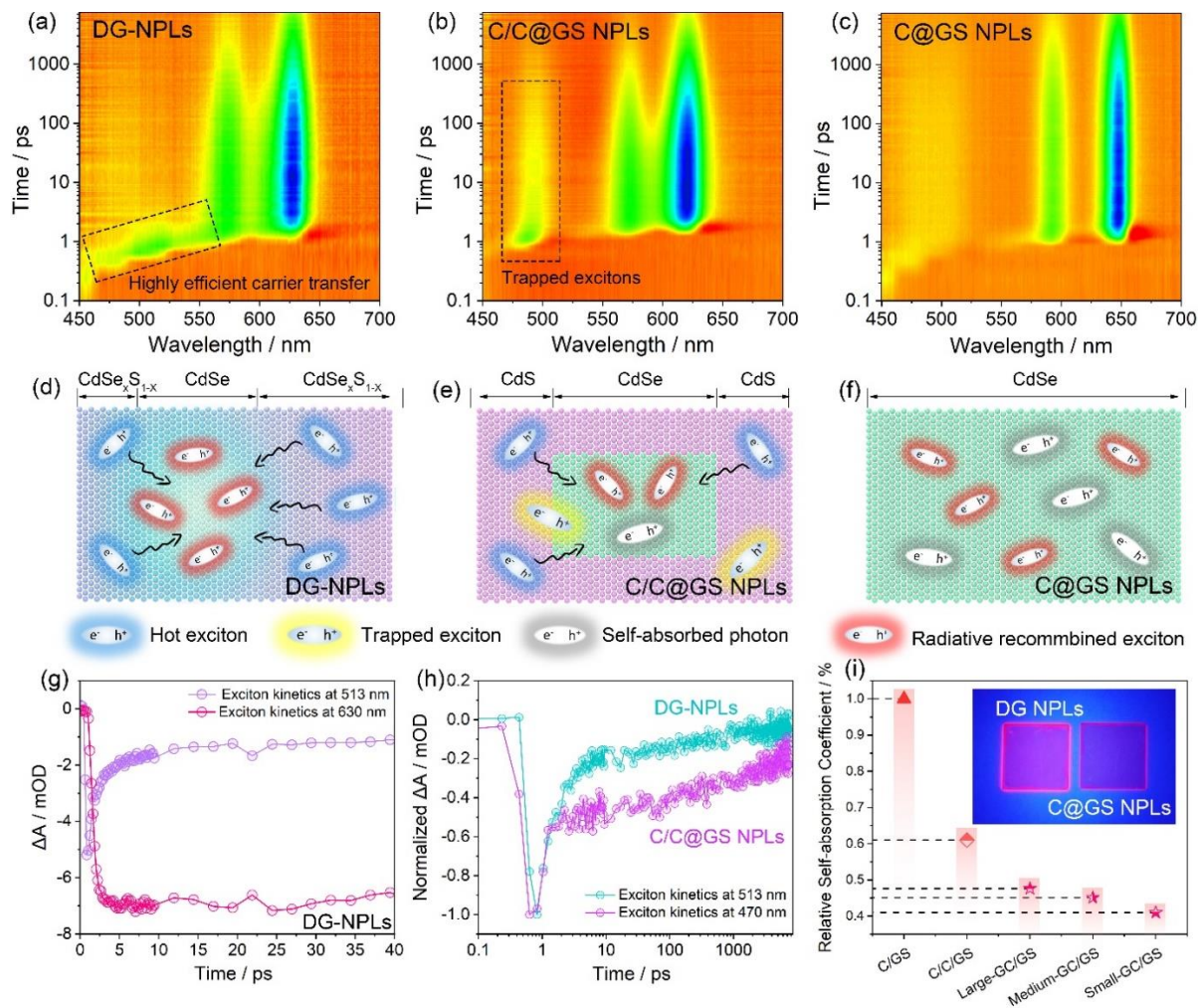

**Figure 2.** Comparative exciton dynamics of DG NPLs, C/C@GS NPLs, and C@GS NPLs. Transient absorption spectroscopies of (a) DG NPLs in comparison to (b) C/C@GS and (e) C@GS NPLs. Schematic illustrations of carriers transfer processes in (d) DG NPLs, (e) C/C@GS NPLs, and (f) C@GS NPLs. (g) Exciton kinetics of DG NPLs at 513 and 630 nm. (h) Exciton kinetics of DG NPLs and C/C@GS NPLs at 513 and 470 nm, respectively. (i) Relative self-absorption coefficient of DG NPLs series in comparison to C@GS and C/C@GS NPLs. Inset shows the photographs of DG NPLs and C@GS NPLs films deposited on quartz substrates under uniform exposure of 405 nm excitation light.

**Figure 1i-k** and **Figure S1d-f** show that gradient cores exhibit tunable photoluminescence (PL) and absorption spectrum, with varying lateral sizes of the CdSe seed. The excitonic peak of the heavy hole continuously blue-shifts with the decrease of CdSe sizes, and the corresponding PL peak shifts from 516 to 511 nm. The atomically flat and uniformly thick surface of all gradient cores ensures a consistent quantum confinement effect in the out-of-plane direction. Therefore, the observed blue-shift in the PL peak is mainly attributed to an additional quantum confinement effect in the in-plane direction, which becomes more prominent with decreasing CdSe seed size.[39, 40] Upon completing the growth of the gradient shell, the DG NPLs series can cover a wider tuning range of 620-645 nm (**Figure 1k**), demonstrating that the in-plane quantum confinement effect becomes more prominent, as the out-of-plane quantum confinement effect relatively weakens, allowing for wider spectral tunability. Notably, all DG NPLs series maintain a near-unity quantum yield and an extremely narrow FWHM of 19-22 nm (**Figure 1l**), attributed to the high-quality shell growth, as evidenced by the TEM and HAADF-STEM images (**Figure S4**), which significantly reduces the inhomogeneous emission broadening by suppressing the core/shell exciton-phonon coupling.[18, 20, 41] It is noteworthy that by reducing the size of CdSe even more, the emission peak can be blue-shifted even further, but it also leads to a broader FWHM of the final product (**Figure S5a**). Additionally, the emission linewidths can also be influenced by changing the Cd/Zn ratio in the shell precursors (**Figure S5b**), which is consistent with previous reports.[20] Moreover, we also synthesized conventional C@GS and core/crown@gradient shell (C/C@GS) structured NPL samples (**Figure S6 & S7**) to compare with the proposed doubly gradient architecture. We find that C@GS NPLs can only achieve near-unity quantum efficiency after sufficient shell growth, losing the ability to control the emission wavelength. Repeating the synthesis of C/C@GS structured NPLs results in a maximum quantum yield of around 85% with an FWHM of approximately 25 nm (**Figure 1l**), consistent with the literature.[33, 38] Therefore, the DG NPLs

demonstrate superior fluorescence properties and wider spectral tunability compared to state-of-the-art C@GS and C/C@GS NPLs.

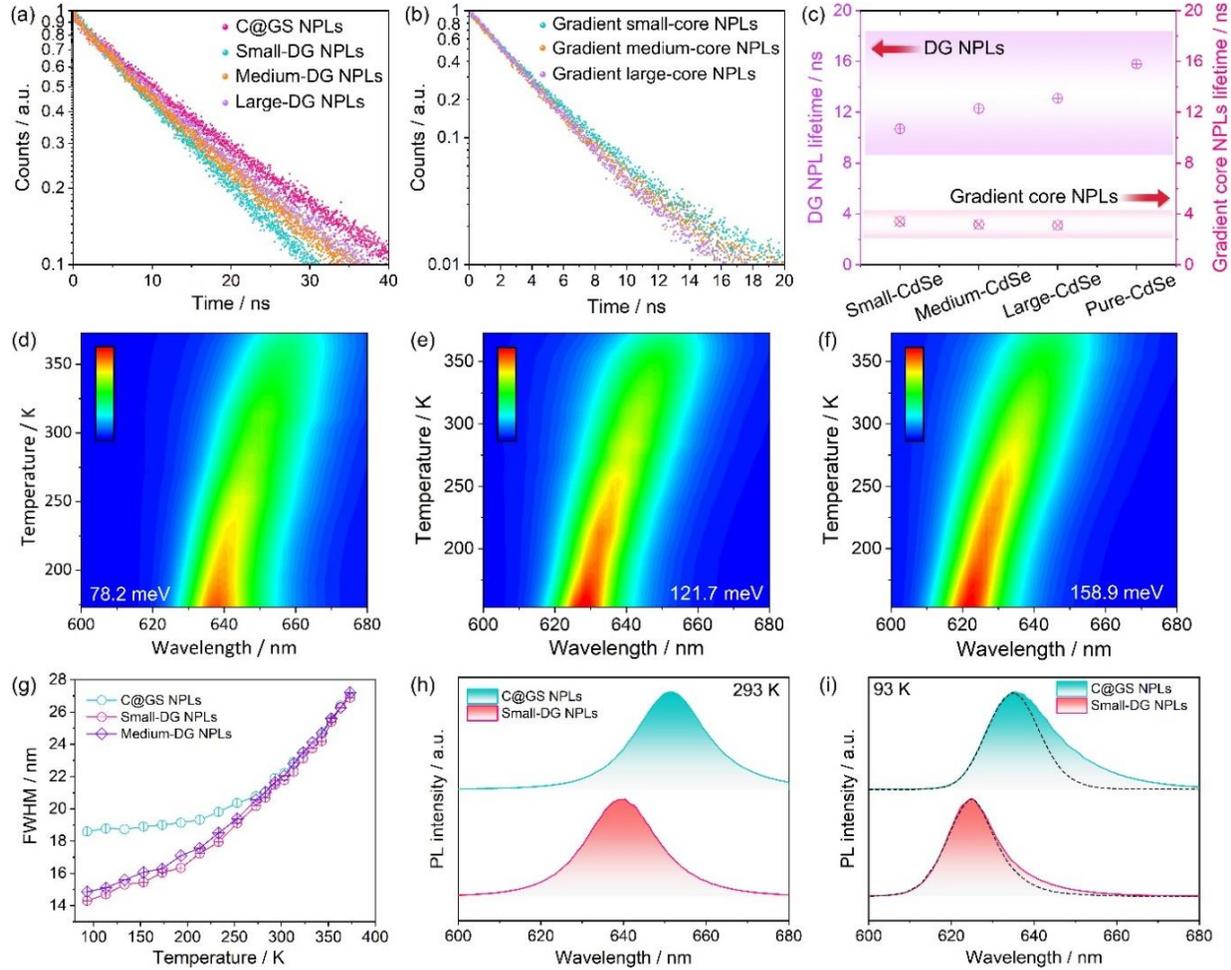

**Figure 3.** Widely tailorable electrostatic interactions in "defect-free" DG NPLs. (a) PL decays of DG NPLs series in comparison to C@GS NPLs. (b) PL decays of gradient cores series. (c) Summarized radiative fluorescent lifetime of DG NPLs, C@GS NPLs and gradient core NPLs. (d-f) 2-D contour plots of temperature-dependent PL spectra of C@GS NPLs and DG NPLs with (e) medium and (f) small size of CdSe seeds, denoted as medium- and small-DG NPLs, respectively. (g) Temperature dependence of FWHMs of small-DG NPLs, medium-DG NPLs, and C@GS NPLs. Comparative PL spectra of small-DG NPLs and C@GS NPLs at (h) 293 K and (i) 93 K, respectively.

To investigate the photophysical properties of DG NPLs in comparison to C@GS and C/C@GS NPLs, we utilized transient absorption (TA) spectroscopy to study their carrier dynamics. Our results, presented in **Figure 2a-c** and **Figure S8**, demonstrate that, upon photoexcitation, DG NPLs exhibit an ultrafast and highly efficient in-plane carrier transfer process, with the bleach signal experiencing a continuous and rapid red-shift from 450 (S-rich region) to 630 nm (CdSe region) in the first ~10 ps (**Figure S9a-c**). Also, during this time, the intensity of the bleach signal in the S-rich region decreases by more than 90%, accompanied by a concomitant increase in the CdSe region (**Figure 2g**). Leveraging the large absorption coefficient of NPLs, the DG NPLs take full advantage of this ultrafast and highly efficient in-plane carrier transfer process to effectively enrich the generated excitons to the R-center (CdSe seed). This step is pivotal in guaranteeing a near-unity quantum yield and validats the expected exceptional exciton enrichment ability of the DG-NPL, as illustrated in **Figure 2d**. Since the DG NPLs series possess similar overall lateral sizes (**Figure S3**), smaller-sized CdSe recombination center will ultimately experience a higher localized exciton concentration *via* this highly efficient carrier transfer process, enabling a large modulation of the strength of electrostatic interactions between electrons and holes. In contrast, the intensity of the bleach signal in the crown region of the C/C@GS NPL only decreases by 50% in the first 10 ps, followed by a much slower decay tail (**Figure 2b & 2h** and **Figure S9d**). Further verifications of the time-resolved PL decay spectra using streak camera (**Figure S10**) demonstrate there were no satellite emissions except for the emission from the CdSe recombination center, indicating that the slow decay process of exciton dynamics in **Figure 2h** can be attributed to the trapped carriers at the core/crown (CdSe/CdS) interface during the relaxation process (**Figure 2e**), resulting in a less efficient quantum yield. Moreover, the in-plane carrier transfer process in DG-NPLs spatially separates exciton generation and recombination (**Figure 2d**), resulting in significantly reduced self-absorption compared to C@GS NPLs, where exciton generation and recombination significantly overlap in space, leading to severe self-absorption (**Figure 2c & 2f**). Notably, the calculated relative self-absorption coefficient indicates that DG NPLs experience up to 59.2% less self-absorption than C@GS NPLs, as summarized in **Figure 2i** (details of the calculation are provided in **Figure S11 & S12**). The corresponding thin films of DG-NPLs and C@GS NPLs (insets in **Figure 2i**) reveal that the DG-NPL film exhibits much brighter edges at the quartz substrate under the same photoexcitation conditions, indicating less energy loss during waveguide propagation due to self-absorption.

By exploiting the inherent carrier transfer mechanism of DG NPLs with type-I band alignments, it is possible to achieve a remarkable level of tunability in the exciton concentration within the R-center by manipulating the relative lateral size of the center compared to the overall NPL, affording a substantial modulation of the strength of electrostatic Coulomb interactions between electrons and holes, which can be deduced by studying their radiative recombination behavior. To compare the radiative behavior of these various exciton-concentration-engineered DG NPLs, we conducted time-resolved photoluminescence (PL) measurements and analyzed their PL decay kinetics in comparison with conventional C@GS NPLs, as presented in **Figure 3a**. All samples display near-mono-exponential PL decay, in excellent agreement with their near-unity quantum yields. However, DG NPLs exhibit a marked reduction in radiative lifetime, dropping from 15.8 (for C@GS NPLs) to 10.2 ns (for small-DG NPLs). This represents a 54.9% increase in the radiative recombination rate, attributed to the strengthened electrostatic Coulomb interactions between electrons and holes within the recombination center, compared to C@GS NPLs. Furthermore, we find that the variations in CdSe size within the gradient core have a relatively small impact on the PL decay kinetics and display an opposite trend to that of the DG-NPL structure (**Figure 3b**).

Specifically, as the CdSe size decreases, the radiative lifetime of the gradient core exhibited an increasing trend. This is primarily due to the strong 1-demensional quantum confinement effect originating from the ultrathin thickness (1.2 nm) of the gradient core squeezes electron wavefunction along the out-of-plane direction and planarly extends it throughout the entire gradient core NPLs, as the conduction band offset between CdSe and $CdSe_xS_{1-x}$ is relatively small, reducing the spatial overlap between the wavefunctions of electrons and holes.[38, 42, 43] As a result, the lifetime increases with decreasing CdSe size in all gradient core series. However, the in-plane delocalization effect of electrons in core-shell structured NPLs is mitigated compared to gradient cores as the quantum confinement of core-shell structured NPLs in the out-of-plane direction is considerably compromised. It is worth noting that the competition between electron delocalization and in-plane spatial confinement also exists within the DG NPL structure. Specifically, as CdSe size is further reduced, a reversed increase in lifetime is observed in the DG NPL (**Figure S13**). However, our experiment results show that the largely relaxed quantum confinement in the out-of-plane direction within the core-shell structure still provides ample room for the engineering of in-plane exciton-concentration to tune their electrostatic interactions and radiative behavior (**Figure 3c**).

To further quantify the strength of electrostatic Coulomb interactions between electrons and holes, we conducted temperature-dependent PL analysis, as depicted in **Figure 3d-f.** By fitting integrated PL intensity as a function of the temperature (**Figure S14**), the calculated $E_b$ for C@GS NPLs, medium-DG NPLs, and small-DG NPLs were 78.2, 121.7, and 158.9 meV, respectively, achieving a remarkable enhancement of the $E_b$ by up to 103.2% in the exciton-concentration-engineered DG NPLs. These findings show the highly tunable electrostatic interactions enabled by the exciton-enrichment strategy, which are crucial for achieving high-performance LED devices and ultra-low-threshold exciton-polariton lasing.[22, 23, 26-28] Furthermore, in the temperature-dependent PL results, we observe that the FWHM of both C@GS NPLs and DG NPLs, as shown in **Figure 3g**, show a consistent trend above ~270 K, ascribed to the stronger exciton-phonon coupling with increasing temperature.[44-46] However, below 270 K, the FWHM of C@GS NPLs decreases at a significantly slower rate compared to DG NPLs. Comparison of the PL spectra of samples at 293 and 93 K in **Figure 3h-i** reveal that C@GS NPLs exhibit more asymmetric PL spectra with a pronounced long tail on the lower photon-energy side at low temperatures. Previous studies have attributed the emission on the low-energy side to trap states with longer PL lifetime, as the thermal energy is insufficient to pass the energy barrier of the traps at lower temperatures.[46] In contrast, emission from the trap states in DG NPLs at low temperatures is greatly suppressed. These results strongly suggest that the doubly gradient architecture with smooth interface transition greatly reduces the defects within the NPLs and yields a closer-to-perfect crystal structure.

As building blocks for practical optoelectronic applications, thermally- and photo-stable nanocrystals are highly desirable for long-term use and commercial deployment.[18, 47, 48] Therefore, we systematically carried out comparative investigations on the thermal-and photo-stability of these DG NPLs. Temperature-dependent of real-time PL intensities in **Figure 4a** shows that the thermally induced exciton dissociation effects in DG NPLs have weakened during heating thanks to their higher $E_b$, resulting in a stronger real-time PL intensity level compared to C@GS NPL. Furthermore, under an inert environment, both DG NPLs and C@GS NPLs can maintain ~90% of their PL intensity after being kept at 280 K for 10 min and then cooled to room temperature (**Figure 4b** and **Figure S15**). In contrast, the PL of gradient cores is almost completely quenched after being subjected to the same treatment at 240 K, highlighting the crucial role played by the inorganic shells in maintaining the stability. Additionally, the PL of C/C@GS NPLs and CdSe/ZnS quantum dots (QDs) decrease by 26% and 43% after being subjected to the same conditions at 280 K, respectively, underscoring the effectiveness of smooth interface transitions in minimizing lattice defects caused by lattice expansion and contraction during heat cycles. DG NPLs also exhibit superior performance compared to C@GS NPLs and gradient cores after long-term heat treatment for 12 h at 353 and 393 K (**Figure 4c** and **Figure S16-S18**), by preserving over 96% and 83% of their initial PL intensity, respectively, which can be attributed to the effective passivation of defects from both the basal plane and edges in the doubly gradient architecture.

In addition to thermal stability, photo-stability is also a critical factor, particularly in color conversion applications where prolonged exposure to blue light is often necessary.[49, 50] Results in **Figure 4d-g** demonstrate DG NPLs can maintain more than 80% of their PL intensity even after 12 h of high-power UV exposure. In contrast, C@GS NPLs and QDs can only maintain 67.3% and 66% of their initial PL intensity, respectively. On the other hand, gradient core NPLs experience a nearly complete quenching after a 12-hour UV exposure. These DG NPLs, with a near-unity quantum yield, significantly reduced self-absorption, and exceptional thermal and photo-stability, exhibit remarkable potential for energy-efficient color conversion applications. As a proof of concept shown in **Figure 4h-j**, we have demonstrated their ability to be patterned into micro-sized (10 μm) arrays with a large thickness of 500 nm of each pixel (**Figure S18**), highlighting the potentials of these DG NPLs as highly effective color converters for ultra-high resolution micro-LED displays.

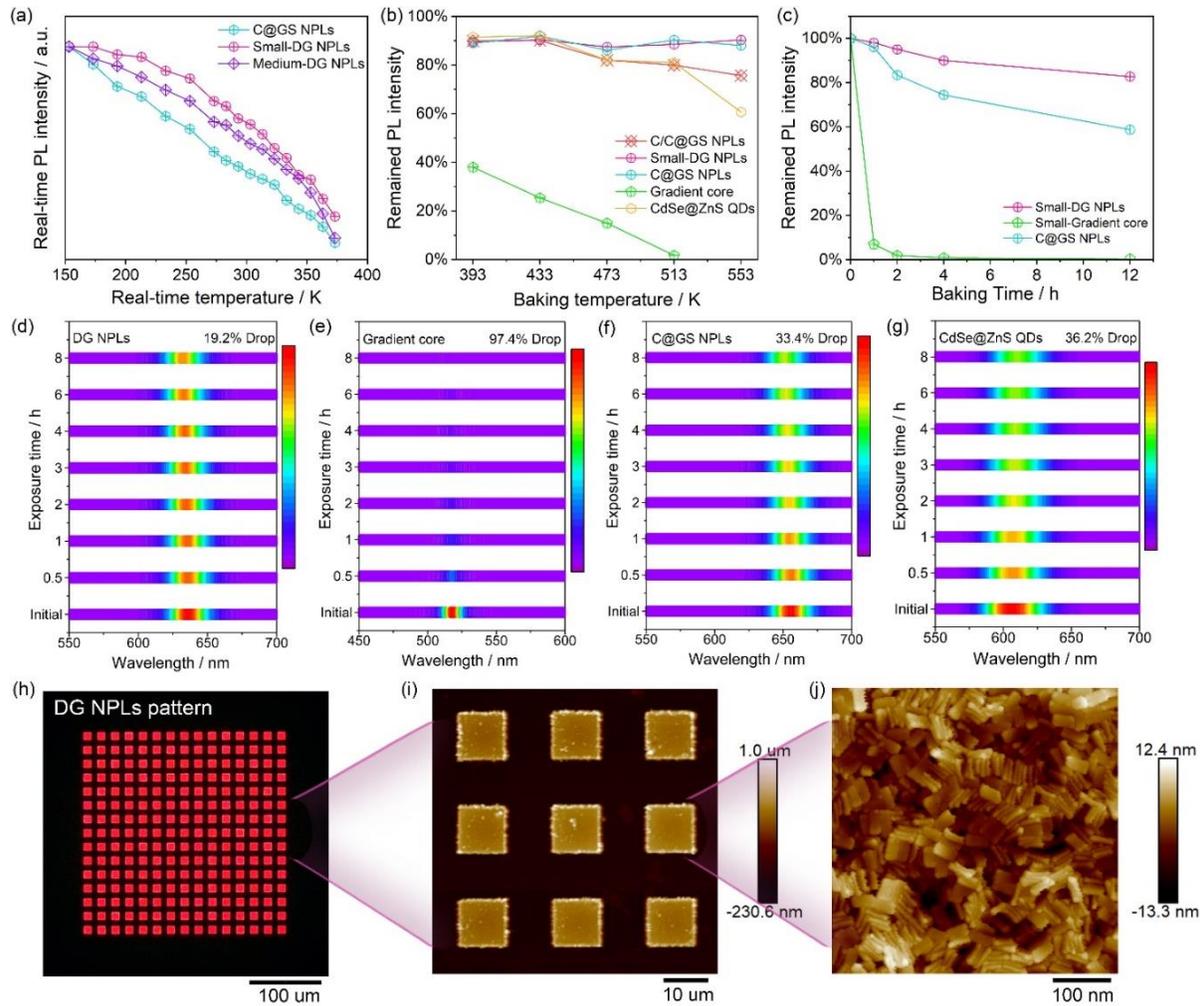

**Figure 4.** Comparative thermal & photo stabilities and conceptual demonstration of DG NPLs as high-resolution micro-color converters. (a) Temperature-dependent of real-time PL intensities of small-DG NPLs, medium-DG NPLs, and C@GS NPLs. (b) Thermal stabilities of structurally different emitters under inert conditions. (c) Long-term thermal stabilities of DG NPLs, gradient core NPLs, and C@GS NPLs at 393 K. Long-term photo stabilities of (d) DG NPLs, (e) gradient core NPLs, (f) C@GS NPLs, and CdSe@ZnS QDs. (h) Fluorescent image of DG NPLs arrays composed of 10 μm-sized pixels. (i) Large-scale and (j) zoomed-in AFM image of DG NPLs pixels.

Furthermore, in LED devices, the appropriate interaction between electrons and holes induced by charge confinement is advantageous for promoting exciton generation and accelerating radiative recombination while suppressing non-radiative Auger recombination, which can be achieved by meticulously tuning the exciton binding energy.[51, 52] Therefore, the highly tunable $E_b$ in exciton-concentration-engineered DG NPLs hold promise for realizing high-performance LED devices. Here, we chose NPLs with different $E_b$ and fabricated three comparative devices using an inverted LED device structure, denoted as small-DG-NPL-LED, medium-DG-NPL-LED, and C@GS-NPL-LED, respectively, as shown in **Figure 5a-b**. Moreover, since the aspect ratio of NPLs plays a crucial role in film smoothness and affects LED device performance,[53] the three different NPLs we used here possess a similar aspect ratio of around 2.3. Results in **Figure 5d-e** show that the medium-DG-NPL-LED exhibits the highest external quantum efficiency of 16.9% and the highest brightness of 43,000 cd/m$^2$, with a 5.7% improvement in EQE and much enhanced maximum brightness compared to C@GS-NPL-LED under the same conditions (current density and luminance variations at different applied voltages are provided in **Figure S19**). Furthermore, this device also exhibits highly stable electroluminescence spectra at different voltages (**Figure 5c**) along with an impressive color purity as evidenced by the CIE coordinates of (0.71,0.29) (**Figure 5f**). Besides, it reaches the peak quantum efficiency at a lower voltage of ~3.3 V, while the C@GS-NPL-LED reaches its peak at 5.0 V (**Figure 5d**). The above results demonstrate that, by appropriately increasing the $E_b$ and enhancing the charge confinements, the medium-DG-NPL-LED achieves a higher carrier recombination efficiency, resulting in overall enhancement in the performance of the LED device. However, for the small-DG-NPL-LED, excessive exciton concentration within the CdSe recombination center accelerates non-radiative Auger recombination of the device under high current density conditions, causing a substantially reduced EQE value. The ability to adjust $E_b$ as needed and the significantly reduced self-absorption coefficient achieved through the DG architecture allows for exciton-concentration-engineered DG NPLs to enhance the performance of LED devices in both EQE and brightness.

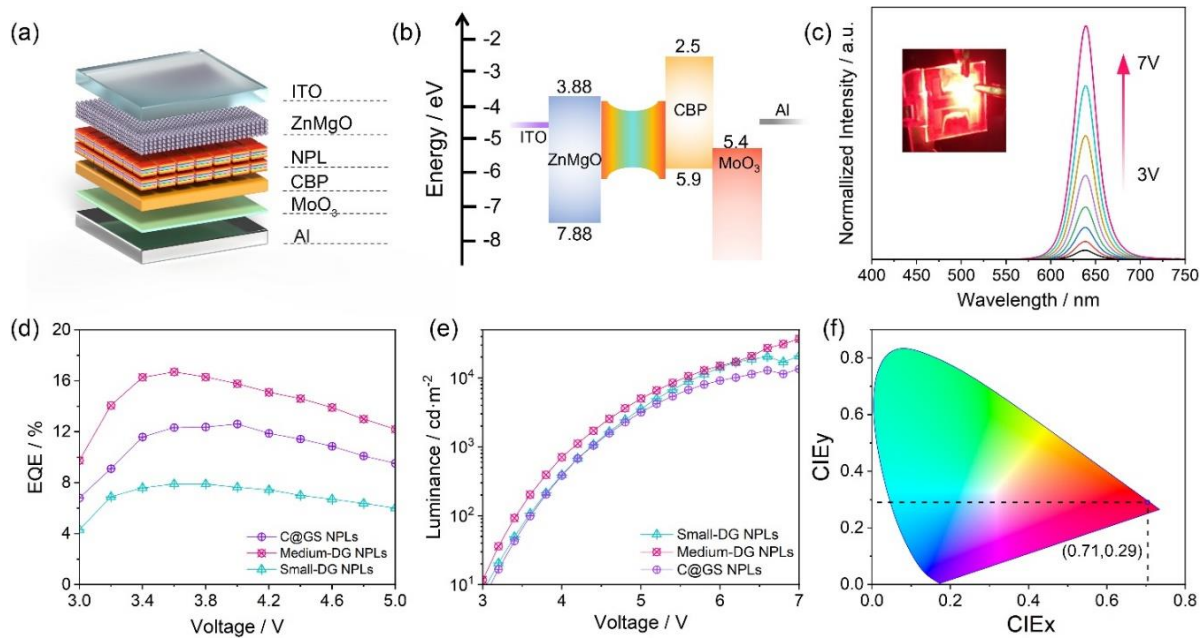

**Figure 5.** High-performance electroluminescence from exciton-concentration-engineered DG NPLs. (a) Schematic illustration of the LED stacking structure. (b) Energy diagram of the medium-DG NPLs-based LEDs. (e) EL spectra of the DG NPLs-based LEDs operating at various applied voltages (d) EQE and (e) luminance of LED devices built from small-DG NPLs, medium-DG NPLs, and C@GS NPLs. (f) Chromaticity diagram and position of the PL signal resulting from the medium-DG NPLs.

CONCLUSIONS

In summary, this study introduces a novel doubly gradient architecture for semiconductor NPLs, offering an on-demand, widely adjustable exciton concentration to regulate EI through in-plane structural engineering as well as an ultrafast and highly efficient carrier transfer process. The resulting DG NPLs show customizable exciton binding energies and radiative recombination rates, with the highest $E_b$ and fastest radiative recombination rates being 103.2% and 54.9% greater than state-of-the-art C@GS NPLs, respectively. Moreover, the spatial separation of exciton generation and recombination in the doubly gradient NPL architecture also leads to a notably reduced self-absorption of up to 59.2%. Furthermore, the DG NPLs exhibit near-unity quantum yields, adjustable emission wavelength from 620 to 645 nm, narrow FWHM in the range of 19-22 nm, and impressive stability. The above flexible features of DG NPLs make them highly customizable for specific optoelectronic applications. As proof-of-concept demonstrations, we present their potential applications as highly efficient pixelated color converters and for enabling high-performance LED devices with an EQE of 16.9% and a maximum luminance of 43,000 cd/m$^2$. This study demonstrates a novel approach to overcoming the impediments of advancing colloidal II-VI semiconductor NPLs for various optoelectronic applications and achieving their high-performance colloidal optoelectronic systems.

METHODS

**Optical characterizations:** The UV-Vis absorption and emission spectra of the NPLs were measured with a Shimadzu UV-1800 spectrophotometer and a Shimadzu RF-5301 PC Spectrofluorophotometer, respectively. To measure the quantum yields, the samples were excited using a 405 nm laser within an integrating sphere, and the data was acquired through an Ocean Optics S4000 spectrometer. To perform time-resolved photoluminescence (TRPL) measurements, a time-correlated single-photon counting (TCSPC) system was employed, utilizing a PicoHarp 300 that can achieve a time resolution as low as 4 ps, which delivered laser pulses at an 80 MHz repetition rate. The setup comprised a driver module (PDL-800 series) that drove a picosecond pulsed laser with a photon energy output of 3.31 eV (375 nm), and a fast photomultiplier tube (Hamamatsu H5783 series) capable of resolving lifetimes at the picosecond level. A streak camera (Optronis) system with a temporal resolution of ~50 ps was also employed for TRPL measurement to resolve the PL spectra evolutions at different decay intervals. To examine the carrier dynamics of the samples, transient absorption (TA) spectroscopy was performed in transmission mode using a Helios setup (Ultrafast Systems LLC) with chirp correction. The pump beam spot size was approximately 50 μm, and the white light continuum probe beam (ranging from 400 to 800 nm) was produced from a 3 mm sapphire crystal by utilizing an 800 nm pulse from the regenerative amplifier. The ultraviolet-visible region detector (CMOS sensor) was utilized to collect the probe beam after it had passed through the sample. Temperature-dependent PL (TDPL) measurement was conducted with a Linkam THMS600 temperature microscope stage. This system allows to heat and freeze the sample in the range from -78 to 873 K. Cooling was achieved using liquid nitrogen, while electrical heating was used for the temperature increase. The sample was measured in the range of 373 K to 153 K with decreasing temperature. The sample was placed in the hermetically sealed chamber and kept under a nitrogen atmosphere to prevent condensation of the water from the air. The stage was assembled with the Witec confocal Raman microscope.

**Structural characterizations and elemental analysis:** For structural and elemental characterizations of the synthesized NPLs, a JEOL TEM 2100F transmission electron microscope was employed, which operated at 200 kV in the high-angle annular dark-field scanning transmission electron microscopy (HAADF-STEM) configuration, equipped with an energy-dispersive X-ray spectroscopy (EDX) detector.

**Atomic force microscopy (AFM) measurement:** The AFM measurement was performed using the scan-assist mode with 512 resolutions by Bruker Dimension Icon scanning probe microscope (Bruker 18 Co., Germany).

**LED device characterization:** A PR705 spectra scan spectrometer was used to record the CIE coordinates and EL spectra. Measurements of the current density and luminance as a function of the applied voltage were performed using a computer-controlled source meter consisting of a programmable Agilent B2902A source meter and a Konica-Minolta LS-110 luminance meter. All measurements were conducted in air at room temperature. The EQE values of the fabricated LED devices were calculated from the obtained luminance, current density, and EL spectrum.

ASSOCIATED CONTENT

**Supporting Information**

Supporting Information is available free of charge at xxxx.

Detailed information about materials synthesis, device fabrication, as well as additional STEM / TEM images, absorption / PL spectra, size distributions, TA spectra, time-resolved PL spectra, calculations of the relative self-absorption coefficient, integrated PL intensity as a function of temperature, stability measurements, AFM characterizations, and applied voltage dependent current density and luminance of the samples.

AUTHOR INFORMATION

**Corresponding Author**

Hilmi Volkan Demir － LUMINOUS! Center of Excellence for Semiconductor Lighting and Displays, The Photonics Institute, School of Electrical and Electronic Engineering, School of Physical and Mathematical Sciences, Nanyang Technological University, Singapore, 639798, UNAM—Institute of Materials Science and Nanotechnology, The National Nanotechnology Research Center, Department of Electrical and Electronics Engineering, Department of Physics, Bilkent University, Bilkent, Ankara, 06800, Turkey; Email: hvdemir@ntu.edu.sg

**Authors**

**Xiao Liang** － LUMINOUS! Center of Excellence for Semiconductor Lighting and Displays, The Photonics Institute, School of Electrical and Electronic Engineering, Nanyang Technological University, Singapore, 639798, Singapore

**Emek G. Durmusoglu** － LUMINOUS! Center of Excellence for Semiconductor Lighting and Displays, The Photonics Institute, School of Electrical and Electronic Engineering, School of Physical and Mathematical Sciences, Nanyang Technological University, Singapore, 639798, Singapore

**Maria Lunina** － Interdisciplinary Graduate Program, Nanyang Technological University, Singapore, 637371 Singapore

**Pedro Ludwig Hernandez-Martinez** － LUMINOUS! Center of Excellence for Semiconductor Lighting and Displays, The Photonics Institute, School of Electrical and Electronic Engineering, Nanyang Technological University, Singapore, 639798, Singapore

**Vytautas Valuckas** － Institute of Materials Research and Engineering, A*STAR (Agency for Science, Technology and Research), 2 Fusionopolis Way, #08-03 Innovis, 138634, Singapore

**Fei Yan** － LUMINOUS! Center of Excellence for Semiconductor Lighting and Displays, The Photonics Institute, School of Electrical and Electronic Engineering, Nanyang Technological University, Singapore, 639798, Singapore

**Yulia Lekina** － Division of Physics and Applied Physics, School of Physical and Mathematical Sciences, Nanyang Technological University, Singapore, 637371, Singapore

**Vijay Kumar Sharma** － LUMINOUS! Center of Excellence for Semiconductor Lighting and Displays, The Photonics Institute, School of Electrical and Electronic Engineering, Nanyang Technological University, Singapore, 639798, Singapore

**Tingting Yin** － Division of Physics and Applied Physics, School of Physical and Mathematical Sciences, Nanyang Technological University, Singapore, 637371, Singapore

**Son Tung Ha** － Institute of Materials Research and Engineering, A*STAR (Agency for Science, Technology and Research), 2 Fusionopolis Way, #08-03 Innovis, 138634, Singapore

**Ze Xiang Shen** － Division of Physics and Applied Physics, School of Physical and Mathematical Sciences, Nanyang Technological University, Singapore, 637371, Singapore

**Handong Sun** － Division of Physics and Applied Physics, School of Physical and Mathematical Sciences, Nanyang Technological University, Singapore, 637371, Singapore

**Arseniy Kuznetsov** － Institute of Materials Research and Engineering, A*STAR (Agency for Science, Technology and Research), 2 Fusionopolis Way, #08-03 Innovis, 138634, Singapore

**Author Contributions**

H.V.D. and X.L. conceived the idea, and H.V.D. supervised the research at all stages. X.L. conducted materials synthesis, device fabrication, steady-state optical spectroscopy, TEM, EDS, TA, TCSPC-based TRPL, AFM, and device characterizations, as well as data analyses. E.D.G. provided technical advice on TA measurements. L.M. and L.Y. carried out TDPL measurements. P.L.H.M. provided simulation support for data analyses. Y.F. and V.K.S. provided technical advice on device fabrication. V.V. fabricated PMMA templates using electron-beam lithography (EBL). T.Y. carried out TRPL measurements using a streak camera, supervised by H.S., H.S.T.

and A.K. supervised the EBL fabrication. Z.X.S. supervised TDPL measurements. X.L. prepared the draft manuscript with input from L.M. and V.V.. H.V.D. and X.L. revised and finalized the manuscript with inputs from all authors.

**Notes**

The authors declare no competing financial interest.

ACKNOWLEDGMENTS

The authors gratefully acknowledge the support from Singapore Agency for Science, Technology and Research (A*STAR) MTC program (Grant No. M21J9b0085) and the Ministry of Education, Singapore (Academic Research Fund Tier 1, MOERG62/20). Partial support was also provided by TUBITAK 119N343, 120N076, 121C266, 121N395, and 20AG001. H.V.D. would like to acknowledge the support received from the TUBA and TUBITAK 2247-A National Leader Researchers Program (121C266). Z.X.S. would like to acknowledge the support from Ministry of Education, Singapore (Tier 1 RG57/21, Tier 2 MOE-T2EP50220-0020 and MOE-T2EP50122-0005). H.S. would like to acknowledge the support from National Research Foundation (NRF-CRP23-2019-0007) and the Ministry of Education, Singapore (Tier 1-RG139/22). Additionally, the authors would like to acknowledge the Facility for Analysis, Characterization, Testing and Simulation (FACTS) at Nanyang Technological University, Singapore, and specifically Dr. Tay Yee Yan and Dr. Andrew Wong, for their valuable and professional technical supports in TEM characterizations. The authors would also like to express their gratitude to the Campus for Research Excellence and Technological Enterprise (CREATE), Singapore, and specifically Prof. Lydia Wong, Dr. Anupam Sadhu, and Mr. Teh Chee Kuang, for sharing the facility and providing technical training for TRPL measurements.

REFERENCES

(1) Passarelli, J. V.; Mauck, C. M.; Winslow, S. W.; Perkinson, C. F.; Bard, J. C.; Sai, H.; Williams, K. W.; Narayanan, A.; Fairfield, D. J.; Hendricks, M. P. Tunable exciton binding energy in 2D hybrid layered perovskites through donor–acceptor interactions within the organic layer. *Nat. Chem.* **2020**, *12* (8), 672-682.

(2) Raja, A.; Chaves, A.; Yu, J.; Arefe, G.; Hill, H. M.; Rigosi, A. F.; Berkelbach, T. C.; Nagler, P.; Schüller, C.; Korn, T. Coulomb engineering of the bandgap and excitons in two-dimensional materials. *Nat. Commun.* **2017**, *8* (1), 15251.

(3) Dong, S.; Puppin, M.; Pincelli, T.; Beaulieu, S.; Christiansen, D.; Hübener, H.; Nicholson, C. W.; Xian, R. P.; Dendzik, M.; Deng, Y. Direct measurement of key exciton properties: Energy, dynamics, and spatial distribution of the wave function. *Nat. Sci.* **2021**, *1* (1), e10010.

(4) Fakharuddin, A.; Gangishetty, M. K.; Abdi-Jalebi, M.; Chin, S.-H.; bin Mohd Yusoff, A. R.; Congreve, D. N.; Tress, W.; Deschler, F.; Vasilopoulou, M.; Bolink, H. J. Perovskite light-emitting diodes. *Nat. Electron.* **2022**, *5* (4), 203-216.

(5) Jang, E.; Jang, H. Quantum Dot Light-Emitting Diodes. *Chem. Rev.* **2023**.

(6) Yang, Z.; Gao, M.; Wu, W.; Yang, X.; Sun, X. W.; Zhang, J.; Wang, H.-C.; Liu, R.-S.; Han, C.-Y.; Yang, H. Recent advances in quantum dot-based light-emitting devices: Challenges and possible solutions. *Mater. Today* **2019**, *24*, 69-93.

(7) Jung, H.; Ahn, N.; Klimov, V. I. Prospects and challenges of colloidal quantum dot laser diodes. *Nat. Photonics* **2021**, *15* (9), 643-655.

(8) Grim, J. Q.; Christodoulou, S.; Di Stasio, F.; Krahne, R.; Cingolani, R.; Manna, L.; Moreels, I. Continuous-wave biexciton lasing at room temperature using solution-processed quantum wells. *Nat. Nanotechnol.* **2014**, *9* (11), 891-895.

(9) Park, Y.-S.; Roh, J.; Diroll, B. T.; Schaller, R. D.; Klimov, V. I. Colloidal quantum dot lasers. *Nat. Rev. Mater.* **2021**, *6* (5), 382-401.

(10) Kagan, C. R.; Bassett, L. C.; Murray, C. B.; Thompson, S. M. Colloidal quantum dots as platforms for quantum information science. *Chem. Rev.* **2020**, *121* (5), 3186-3233.

(11) García de Arquer, F. P.; Talapin, D. V.; Klimov, V. I.; Arakawa, Y.; Bayer, M.; Sargent, E. H. Semiconductor quantum dots: Technological progress and future challenges. *Science* **2021**, *373* (6555), eaaz8541.

(12) Diroll, B. T.; Guzelturk, B.; Po, H.; Dabard, C.; Fu, N.; Makke, L.; Lhuillier, E.; Ithurria, S. 2D II–VI Semiconductor Nanoplatelets: From Material Synthesis to Optoelectronic Integration. *Chem. Rev.* **2023**.

(13) Bai, B.; Zhang, C.; Dou, Y.; Kong, L.; Wang, L.; Wang, S.; Li, J.; Zhou, Y.; Liu, L.; Liu, B. Atomically flat semiconductor nanoplatelets for light-emitting applications. *Chem. Soc. Rev.* **2023**.

(14) Naeem, A.; Masia, F.; Christodoulou, S.; Moreels, I.; Borri, P.; Langbein, W. Giant exciton oscillator strength and radiatively limited dephasing in two-dimensional platelets. *Phys. Rev. B* **2015**, *91* (12), 121302.

(15) Li, Q.; Lian, T. Exciton spatial coherence and optical gain in colloidal two-dimensional cadmium chalcogenide nanoplatelets. *Acc. Chem. Res.* **2019**, *52* (9), 2684-2693.

(16) Rowland, C. E.; Fedin, I.; Diroll, B. T.; Liu, Y.; Talapin, D. V.; Schaller, R. D. Elevated temperature photophysical properties and morphological stability of CdSe and CdSe/CdS nanoplatelets. *J. Phys. Chem. Lett.* **2018**, *9* (2), 286-293.

(17) Sharma, M.; Delikanli, S.; Demir, H. V. Two-dimensional CdSe-based nanoplatelets: their heterostructures, doping, photophysical properties, and applications. *Proc. IEEE* **2019**, *108* (5), 655-675.

(18) Altintas, Y.; Quliyeva, U.; Gungor, K.; Erdem, O.; Kelestemur, Y.; Mutlugun, E.; Kovalenko, M. V.; Demir, H. V. Highly Stable, Near-Unity Efficiency Atomically Flat Semiconductor Nanocrystals of CdSe/ZnS Hetero-Nanoplatelets Enabled by ZnS-Shell Hot-Injection Growth. *Small* **2019**, *15* (8), 1804854.

(19) Davis, A. H.; Hofman, E.; Chen, K.; Li, Z.-J.; Khammang, A.; Zamani, H.; Franck, J. M.; Maye, M. M.; Meulenberg, R. W.; Zheng, W. Exciton energy shifts and tunable dopant emission in manganese-doped two-dimensional CdS/ZnS core/shell nanoplatelets. *Chem. Mater.* **2019**, *31* (7), 2516-2523.

(20) Altintas, Y.; Gungor, K.; Gao, Y.; Sak, M.; Quliyeva, U.; Bappi, G.; Mutlugun, E.; Sargent, E. H.; Demir, H. V. Giant alloyed hot injection shells enable ultralow optical gain threshold in colloidal quantum wells. *ACS Nano* **2019**, *13* (9), 10662-10670.

(21) Hazarika, A.; Fedin, I.; Hong, L.; Guo, J.; Srivastava, V.; Cho, W.; Coropceanu, I.; Portner, J.; Diroll, B. T.; Philbin, J. P. Colloidal atomic layer deposition with stationary reactant phases enables precise synthesis of "digital" II–VI nano-heterostructures with exquisite control of confinement and strain. *J. Am. Chem. Soc.* **2019**, *141* (34), 13487-13496.

(22) Deng, Y.; Lin, X.; Fang, W.; Di, D.; Wang, L.; Friend, R. H.; Peng, X.; Jin, Y. Deciphering exciton-generation processes in quantum-dot electroluminescence. *Nat. Commun.* **2020**, *11* (1), 2309.

(23) Kim, J. S.; Heo, J.-M.; Park, G.-S.; Woo, S.-J.; Cho, C.; Yun, H. J.; Kim, D.-H.; Park, J.; Lee, S.-C.; Park, S.-H. Ultra-bright, efficient and stable perovskite light-emitting diodes. *Nature* **2022**, 1-7.

(24) Qu, J.; Rastogi, P.; Gréboval, C.; Livache, C.; Dufour, M.; Chu, A.; Chee, S.-S.; Ramade, J.; Xu, X. Z.; Ithurria, S. Nanoplatelet-based light-emitting diode and its use in all-nanocrystal LiFi-like communication. *ACS Appl. Mater. Inter.* **2020**, *12* (19), 22058-22065.

(25) Turtos, R.; Gundacker, S.; Polovitsyn, A.; Christodoulou, S.; Salomoni, M.; Auffray, E.; Moreels, I.; Lecoq, P.; Grim, J. Ultrafast emission from colloidal nanocrystals under pulsed X-ray excitation. *J. Instrum.* **2016**, *11* (10), P10015.

(26) Su, R.; Fieramosca, A.; Zhang, Q.; Nguyen, H. S.; Deleporte, E.; Chen, Z.; Sanvitto, D.; Liew, T. C.; Xiong, Q. Perovskite semiconductors for room-temperature exciton-polaritonics. *Nat. Mater.* **2021**, *20* (10), 1315-1324.

(27) Fraser, M. D.; Höfling, S.; Yamamoto, Y. Physics and applications of exciton–polariton lasers. *Nat. Mater.* **2016**, *15* (10), 1049-1052.

(28) Du, W.; Zhang, S.; Zhang, Q.; Liu, X. Recent progress of strong exciton–photon coupling in lead halide perovskites. *Adv. Mater.* **2019**, *31* (45), 1804894.

(29) Sharma, M.; Gungor, K.; Yeltik, A.; Olutas, M.; Guzelturk, B.; Kelestemur, Y.; Erdem, T.; Delikanli, S.; McBride, J. R.; Demir, H. V. Near-unity emitting copper-doped colloidal semiconductor quantum wells for luminescent solar concentrators. *Adv. Mater.* **2017**, *29* (30), 1700821.

(30) Zhang, J.; Sun, Y.; Ye, S.; Song, J.; Qu, J. Heterostructures in two-dimensional CdSe nanoplatelets: synthesis, optical properties, and applications. *Chem. Mater.* **2020**, *32* (22), 9490-9507.

(31) Li, Q.; Wu, K.; Chen, J.; Chen, Z.; McBride, J. R.; Lian, T. Size-independent exciton localization efficiency in colloidal CdSe/CdS core/crown nanosheet type-I heterostructures. *ACS Nano* **2016**, *10* (3), 3843-3851.

(32) Kunneman, L. T.; Schins, J. M.; Pedetti, S.; Heuclin, H.; Grozema, F. C.; Houtepen, A. J.; Dubertret, B.; Siebbeles, L. D. Nature and decay pathways of photoexcited states in CdSe and CdSe/CdS nanoplatelets. *Nano Lett.* **2014**, *14* (12), 7039-7045.

(33) Shabani, F.; Dehghanpour Baruj, H.; Yurdakul, I.; Delikanli, S.; Gheshlaghi, N.; Isik, F.; Liu, B.; Altintas, Y.; Canımkurbey, B.; Demir, H. V. Deep-Red-Emitting Colloidal Quantum Well Light-Emitting Diodes Enabled through a Complex Design of Core/Crown/Double Shell Heterostructure. *Small* **2022**, *18* (8), 2106115.

(34) Schlosser, A.; Graf, R. T.; Bigall, N. C. CdS crown growth on CdSe nanoplatelets: core shape matters. *Nanoscale Adv.* **2020**, *2* (10), 4604-4614.

(35) Wen, Z.; Zhang, C.; Zhou, Z.; Xu, B.; Wang, K.; Teo, K. L.; Sun, X. W. Ultrapure green light-emitting diodes based on CdSe/CdS core/crown nanoplatelets. *IEEE J. Quantum Electron.* **2019**, *56* (1), 1-6.

(36) Bae, W. K.; Char, K.; Hur, H.; Lee, S. Single-step synthesis of quantum dots with chemical composition gradients. *Chem. Mater.* **2008**, *20* (2), 531-539.

(37) Rossinelli, A. A.; Rojo, H.; Mule, A. S.; Aellen, M.; Cocina, A.; De Leo, E.; Schäublin, R.; Norris, D. J. Compositional grading for efficient and narrowband emission in cdse-based core/shell nanoplatelets. *Chem. Mater.* **2019**, *31* (22), 9567-9578.

(38) Taghipour, N.; Delikanli, S.; Shendre, S.; Sak, M.; Li, M.; Isik, F.; Tanriover, I.; Guzelturk, B.; Sum, T. C.; Demir, H. V. Sub-single exciton optical gain threshold in colloidal semiconductor quantum wells with gradient alloy shelling. *Nat. Commun.* **2020**, *11* (1), 3305.

(39) Di Giacomo, A.; Rodà, C.; Khan, A. H.; Moreels, I. Colloidal synthesis of laterally confined blue-emitting 3.5 monolayer CdSe nanoplatelets. *Chem. Mater.* **2020**, *32* (21), 9260-9267.

(40) Bertrand, G. H.; Polovitsyn, A.; Christodoulou, S.; Khan, A. H.; Moreels, I. Shape control of zincblende CdSe nanoplatelets. *ChemComm* **2016**, *52* (80), 11975-11978.

(41) Rossinelli, A. A.; Riedinger, A.; Marqués-Gallego, P.; Knüsel, P. N.; Antolinez, F. V.; Norris, D. J. High-temperature growth of thick-shell CdSe/CdS core/shell nanoplatelets. *ChemComm* **2017**, *53* (71), 9938-9941.

(42) Meinardi, F.; Colombo, A.; Velizhanin, K. A.; Simonutti, R.; Lorenzon, M.; Beverina, L.; Viswanatha, R.; Klimov, V. I.; Brovelli, S. Large-area luminescent solar concentrators based on 'Stokes-shift-engineered' nanocrystals in a mass-polymerized PMMA matrix. *Nat. Photonics* **2014**, *8* (5), 392-399.

(43) Wang, C.; Barba, D.; Selopal, G. S.; Zhao, H.; Liu, J.; Zhang, H.; Sun, S.; Rosei, F. Enhanced photocurrent generation in proton-irradiated "giant" CdSe/CdS core/shell quantum dots. *Adv. Funct. Mater.* **2019**, *29* (46), 1904501.

(44) Liu, X.; Zhang, X.; Li, L.; Xu, J.; Yu, S.; Gong, X.; Zhang, J.; Yin, H. Stable Luminescence of CsPbBr3/n CdS Core/Shell Perovskite Quantum Dots with Al Self-Passivation Layer Modification. *ACS Appl. Mater. Inter.* **2019**, *11* (43), 40923-40931.

(45) Cheng, O. H.-C.; Qiao, T.; Sheldon, M.; Son, D. H. Size-and temperature-dependent photoluminescence spectra of strongly confined $CsPbBr_3$ quantum dots. *Nanoscale* **2020**, *12* (24), 13113-13118.

(46) Tessier, M.; Mahler, B.; Nadal, B.; Heuclin, H.; Pedetti, S.; Dubertret, B. Spectroscopy of colloidal semiconductor core/shell nanoplatelets with high quantum yield. *Nano Lett.* **2013**, *13* (7), 3321-3328.

(47) Moon, H.; Lee, C.; Lee, W.; Kim, J.; Chae, H. Stability of quantum dots, quantum dot films, and quantum dot light-emitting diodes for display applications. *Adv. Mater.* **2019**, *31* (34), 1804294.

(48) Ghosh, Y.; Mangum, B. D.; Casson, J. L.; Williams, D. J.; Htoon, H.; Hollingsworth, J. A. New insights into the complexities of shell growth and the strong influence of particle volume in nonblinking "giant" core/shell nanocrystal quantum dots. *J. Am. Chem. Soc.* **2012**, *134* (23), 9634-9643.

(49) Kim, O.-H.; Ha, S.-W.; Kim, J. I.; Lee, J.-K. Excellent photostability of phosphorescent nanoparticles and their application as a color converter in light emitting diodes. *ACS Nano* **2010**, *4* (6), 3397-3405.

(50) Jun, S.; Lee, J.; Jang, E. Highly luminescent and photostable quantum dot–silica monolith and its application to light-emitting diodes. *ACS Nano* **2013**, *7* (2), 1472-1477.

(51) Yang, X.; Zhang, X.; Deng, J.; Chu, Z.; Jiang, Q.; Meng, J.; Wang, P.; Zhang, L.; Yin, Z.; You, J. Efficient green light-emitting diodes based on quasi-two-dimensional composition and phase engineered perovskite with surface passivation. *Nat. Commun.* **2018**, *9* (1), 570.

(52) Kumar, S.; Jagielski, J.; Kallikounis, N.; Kim, Y.-H.; Wolf, C.; Jenny, F.; Tian, T.; Hofer, C. J.; Chiu, Y.-C.; Stark, W. J. Ultrapure green light-emitting diodes using two-dimensional formamidinium perovskites: achieving recommendation 2020 color coordinates. *Nano Lett.* **2017**, *17* (9), 5277-5284.

(53) Liu, B.; Altintas, Y.; Wang, L.; Shendre, S.; Sharma, M.; Sun, H.; Mutlugun, E.; Demir, H. V. Record High External Quantum Efficiency of 19.2% Achieved in Light-Emitting Diodes of Colloidal Quantum Wells Enabled by Hot-Injection Shell Growth. *Adv. Mater.* **2020**, *32* (8), 1905824.

# Supplementary Information: Near-Unity Emitting, Widely Tailorable and Stable Exciton Concentrators Built from Doubly Gradient 2D Semiconductor Nanoplatelets

Xiao Liang,[1] Emek G. Durmusoglu,[1,2] Maria Lunina,[3] Pedro Ludwig Hernandez-Martinez,[1] Vytautas Valuckas,[4] Fei Yan,[1] Yulia Lekina,[2] Vijay Kumar Sharma,[1] Tingting Yin,[2] Son Tung Ha,[4] Ze Xiang Shen,[2] Handong Sun,[2] Arseniy Kuznetsov,[4] and Hilmi Volkan Demir[1,2,5]*

[1]LUMINOUS! Center of Excellence for Semiconductor Lighting and Displays, The Photonics Institute, School of Electrical and Electronic Engineering, Nanyang Technological University, Singapore, 639798, Singapore

[2]Division of Physics and Applied Physics, School of Physical and Mathematical Sciences, Nanyang Technological University, Singapore, 637371, Singapore

[3]Interdisciplinary Graduate Program, Nanyang Technological University, Singapore, 637371 Singapore

[4]Institute of Materials Research and Engineering, A*STAR (Agency for Science, Technology and Research), 2 Fusionopolis Way, #08-03 Innovis, 138634, Singapore

[5]UNAM—Institute of Materials Science and Nanotechnology, The National Nanotechnology Research Center, Department of Electrical and Electronics Engineering, Department of Physics, Bilkent University, Bilkent, Ankara, 06800, Turkey

E-mail: hvdemir@ntu.edu.sg

*Chemicals*

Cadmium oxide (CdO, 99.9%), cadmium acetate dihydrate (Cd(OAc)$_2$·2H$_2$O, >98%), cadmium nitrate tetrahydrate (Cd(NO$_3$)$_2$·4H$_2$O, 99.997%, trace metals basis), sodium myristate (≥99%), selenium (Se, 99.99%, trace metals basis), sulfur (S, 99,998%, trace metals basis), zinc acetate (Zn(OAc)$_2$, 99.99%, trace metals basis), znic acetate dihydrate (Zn(OAc)$_2$·2H$_2$O, 99.999%, trace metal basis), 1-octanethiol (≥98.5%), trioctylamine (TOA, 98%), magnesium acetate tetrahydrate (Mg(OAc)$_2$·4H$_2$O, ≥99%), tetramethylammonium hydroxide pentahydrate (TMAH, ≥97%), ethanolamine (≥99%), dimethyl sulfoxide (DMSO, ≥99.9%) 1-octadecene (ODE, technical-grade), oleic acid (OA, 90%), oleylamine (OAm, 70%), and trioctylphoshine (TOP, 90%) were purchased from Sigma-Aldrich. Methanol, hexane, butanol, acetone, toluene, and ethanol were obtained from Merck Millipore and used without any further purification.

*Preparation of Cadmium myristate precursors*

The synthesis of cadmium myristate (Cd(Myr)$_2$) was carried out by following a previously reported method. [1] Firstly, 1.23 g of Cd(NO$_3$)$_2$·4H$_2$O and 3.13 g of sodium myristate was separately dissolved in 40 mL and 250 mL of methanol, respectively. Once both chemicals had completely dissolved, the solutions were mixed and stirred vigorously for 1 h, resulting in the precipitation of Cd(Myr)$_2$. The precipitate was separated by centrifugation and then washed by redispersing in methanol to remove any excess or unreacted precursors. The above washing step was repeated at least three times, and the precipitate was dried at 50 °C under vacuum overnight before use.

*Preparation of 0.1 M S-ODE precursors*

0.1 M S-ODE solution was prepared by fully dissolving 32 mg of S in 10 mL of ODE using ultrasonic for 30 min.

*Preparation of anisotropic growth solution for CdS crown*

A mixture of 480 mg of Cd(OAc)$_2$·2H$_2$O, 340 μL of OA, and 2 mL of ODE in a 10 mL three-neck flask was sonicated for 30 min at room temperature. The solution was then heated to 160 °C under continuous stirring in ambient atmosphere until the formation of a whitish colored homogeneous gel. After that, the above cadmium precursors were combined with 3 mL of 0.1 M S-ODE precursors for the CdS crown growth.

*Preparation of 0.25 M Zn(OA)$_2$ precursors*

0.92 g of Zn(OAc)$_2$ and 2.8 g of OA were mixed with 20 mL of TOA in a 100 mL three-neck flask. The above mixture was then degassed under room temperature for 30 min. Subsequently, the temperature was heated to 200 °C for 20 min under N$_2$ protection until the mixture became clear. The resulting solution was heated to 100 °C before use.

*Synthesis of CdSe/CdSe$_x$S$_{1-x}$ 4ML Gradient cores*

Typically, a mixture of 170 mg of Cd(Myr)$_2$, 12 mg of Se powder, and 15 mL of ODE was introduced into a 50 mL three-neck flask. After 30 min of degassing at room temperature, the atmosphere was replaced with N$_2$ and the temperature was set to 230 °C. At approximately 195 °C, 80 mg of Cd(OAc)$_2$·2H$_2$O was rapidly added, and the gradient transition from CdSe to CdS was initiated by introducing a total 1.5 mL of 0.1 M S-ODE precursors at a steady rate into the same flask after the CdSe seeds had reached a certain size. The resulting solution was heated at 240 °C for 8 min before being cooled to room temperature. At 150 °C, 1 mL of OA was introduced, followed by the addition of 10 mL of hexane at 60°C. The resulting gradient cores were then precipitated by adding 5 mL of ethanol and separated by centrifugating at 4000 rpm for 10 min. Finally, the as-made gradient cores were washed once more before being ultimately redispersed in 5 mL of hexane for further use.

*Synthesis of quasi-squarish and quasi-rectangular 4ML CdSe cores*

4ML CdSe cores with quasi-squarish shape were synthesized in accordance with established protocols with slight modifications.[2] Specifically, 170 mg of Cd(Myr)$_2$ and 12 mg of Se powder were mixed with 15 mL of ODE in a 50 mL three-necked flask. Following a 20 min degassing period at room temperature, the solution was purged with N$_2$ and heated to 240 °C. At approximately 195 °C, 80 mg of Cd(OAc)$_2$·2H$_2$O was swiftly added. After 10 min growth, the solution was cooled to room temperature. During the cooling process, 1 mL of OA was introduced at 150 °C, followed by the addition of 10 mL of hexane at 60°C. The resulting CdSe 4ML cores were precipitated by introducing 5 mL of ethanol and separated by centrifugating at 4000 rpm for 10 min. Finally, the as-made NPLs were washed once more before being ultimately redispersed in 5 mL of hexane for further use. For the synthesis of CdSe 4ML core NPLs possessing a quasi-rectangular shape (aspect ratio of ~2), Cd(OAc)$_2$·2H$_2$O was substituted with dried Cd(OAc)$_2$, with all remaining procedures identical.

*Synthesis of CdSe/CdS 4ML core/crown NPLs*

CdSe/CdS 4ML core/crown NPLs were synthesized according to an establish protocol.[3] Typically, 1 mL of 4ML CdSe hexane dispersion, along with 5 mL of ODE and 100 μL of OA, was mixed within a 50 mL three-neck flask. The above solution was then degassed at 80 °C to fully eliminate hexane and air. Subsequently, the temperature was raised to 240 °C under N$_2$ protection. After that, the anisotropic CdS crown growth mixture was slowly introduced into the flask at a rate of 8 mL/h. Then, the resultant mixture was further annealed for 5 min at 240 °C. The solution was then cooled to room temperature, and the CdSe/CdS core/crown products were precipitated by using ethanol. Ultimately, the precipitated NPLs were dissolved in hexane for further use.

*Gradient shell growth for synthesizing DG NPLs, C/C@GS NPLs, and C@GS NPLs*

A mixture of 23 mg of $Cd(OAc)_2 \cdot 2H_2O$, 55 mg of $Zn(OAc)_2$, 1 mL of OA, and 10 mL of ODE was combined in a 50 mL three-neck flask, which was subjected to vacuum at room temperature for 20 min, followed by switching the atmosphere to $N_2$. Subsequently, the mixture was heated to 200 °C and maintained at this temperature for 30 min until a clear solution was obtained. The temperature was then lowered to 60 °C, and either 1 mL of $CdSe/CdSe_xS_{1-x}$ 4ML gradient cores, CdSe/CdS 4ML core/crown NPLs or CdSe 4ML cores in hexane was introduced. Then, the mixture was degassed for 45 min under vacuum to eliminate residual hexane. After that, $N_2$ flow was initiated through the Schlenk line. The solution was then heated to 300 °C, and 1 mL of degassed OAm was added at 90 °C. A solution of 4 mL of 0.1 M 1-octanethiol in ODE was employed as the S-source, and the injection was carried out at a rate of 8 mL/h using a syringe pump, starting at 165 °C and continuing until the precursor was fully injected. After that, the solution was maintained at this temperature for 40 min of annealing, before the reaction was quenched using a water bath. At 60 °C, 10 mL of hexane was added, and the as-synthesized NPLs were precipitated using 10 mL of ethanol and washed twice. Finally, the precipitated NPLs were redispersed in hexane or toluene for further use.

*Synthesis of CdSe@ZnS QDs*

CdSe@ZnS QDs were synthesized with slight modifications to a previously reported method. [4] Specifically, a mixture of 0.206 g of CdO and 1.8 g of OA was dissolved into 40 mL of TOA and then degassed. The mixture was heated to 150 °C with rapid stirring, followed by further heating to 300 °C under $N_2$ flow. At 300 °C, 0.2 mL of 2.0 M Se-TOP was rapidly injected. After 90 s, 1.2 mmol of 1-octanethiol dissolved in TOA (210 μL in 6 mL) was injected at a rate of 1 mL/min using a syringe pump, and the reaction was allowed to proceed for 40 min. Subsequently, the 0.25 M Zn $(OA)_2$ precursor solution was injected into the Cd-containing reaction medium at a rate of 2 mL/min. After that, 6.4 mmol of 1-octanethiol dissolved in TOA (1.12 mL in 6 mL) was added at a rate of 1 mL/min using a syringe pump. The reaction was allowed to proceed for a total of 2 h. After completion of the reaction, the solution was cooled to room temperature and the organic sludge was removed *via* centrifugation at 5000 rpm for 10 min. Ethanol was then added to the product solution until an opaque flocculant appeared, and the resulting QDs were separated by further centrifugation. Finally, the precipitates were dispersed in toluene/hexane for further use.

### *Synthesis of $Zn_{0.95}Mg_{0.05}O$ nanocrystals*

$Zn_{0.95}Mg_{0.05}O$ nanocrystals were synthesized using a previously reported protocol. [5] Initially, 2.85 mol of $Zn(OAc)_2 \cdot 2H_2O$ and 0.15 mol of $Mg(OAc)_2 \cdot 4H_2O$ were dissolved in 30 mL of DMSO. Next, 1g of TMAH was dissolved in 10 mL of ethanol, and the resulting solution was dropwisely introduced to the above Zn/Mg solution and stirred for 1 h under ambient conditions. After completion of the injection, the solution was stirred at ambient temperature for another 1 h. Subsequently, 250 μL of ethanolamine was added to the solution, and the mixture was stirred for an additional 30 min. Excessive ethyl acetate or acetone were added until the solution became turbid. Finally, the resulting precipitates were separated by centrifuging at 6000 rpm for 3 min and then redispersed in 15 mL of butanol for further use.

*Fabrication of florescent NPLs pattern*

A direct deep patterning technique was utilized to create fluorescent NPLs patterns with slight modifications. [6] Basically, a sacrificial layer of PMMA (Poly(methylmethacrylate)) resist was firstly spin-coated on a quartz slide at 2400 rpm and baked on a hotplate at 180 °C for 3 min, resulting in a 400 nm PMMA film. Subsequently, a thin layer of electrification dissipating material (ESPACER 300AX01) was spin-coated on top of the resist at 1500 rpm. The resulting sample was then patterned using e-beam lithography (Elionix ELS-7000) with acceleration voltage of 100 kV and probe current of 500 pA, developed in MIBK:IPA 1:3 solution for 70 s. After that, the NPLs were deposited onto the substrate with the patterned PMMA sacrificial layer *via* spin-coating. Finally, the PMMA sacrificial layer was lifted-off by ultrasonication in acetone for 20 s, leaving only the developed NPL patterns on the quartz substrate.

*LED device fabrication*

$Zn_{0.95}Mg_{0.05}O$ nanocrystals, used as electron transport layer, were deposited on a ITO-coated glass substrate *via* spin-coating, followed by baking at 120 °C for 30 min in the glovebox. After that, NPLs were deposited onto the above substrate by spin-coating at 2000 rpm for 60 s. Subsequently, the resulting substrate were transferred to a vacuum thermal evaporation chamber for the deposition of CBP, $MoO_3$ and Al layers at a base pressure of $2.5 \times 10^{-4}$ Pa, with the thicknesses of each layer being accurately controlled using quartz crystal oscillators. Finally, under $N_2$ atmosphere, the devices were encapsulated immediately using epoxy glue and glass slides.

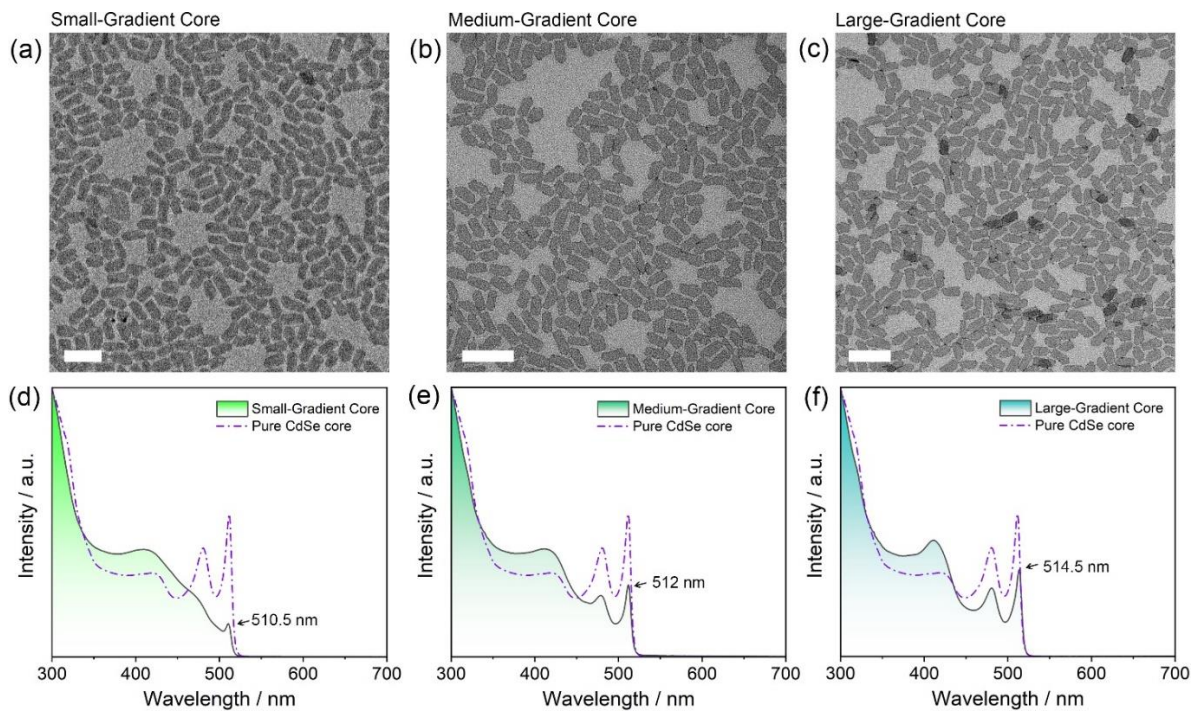

**Figure S1.** TEM images of the (a) small-, (b) medium-, and (c) large-gradient cores (scale bar: 50 nm). Steady-state absorption spectra of the (d) small-, (e) medium-, and (f) large-gradient cores with the start injection time of S precursors at 100, 200, and 300 s, respectively. The dash-dotted lines represent the normalized absorption spectra of the pure 4ML CdSe cores. It can be seen that the earlier injection of S precursors leads to a gradual decrease in the relative intensity of the excitonic absorption peak of CdSe. This observation indicates a reduction in the area of CdSe recombination center relative to the entire gradient core area, which is consistent with the results obtained from the EDS line-scan, demonstrating that the relative proportion of CdSe (recombination center) to the entire gradient core can be effectively controlled by regulating the initiation injection time of S precursors.

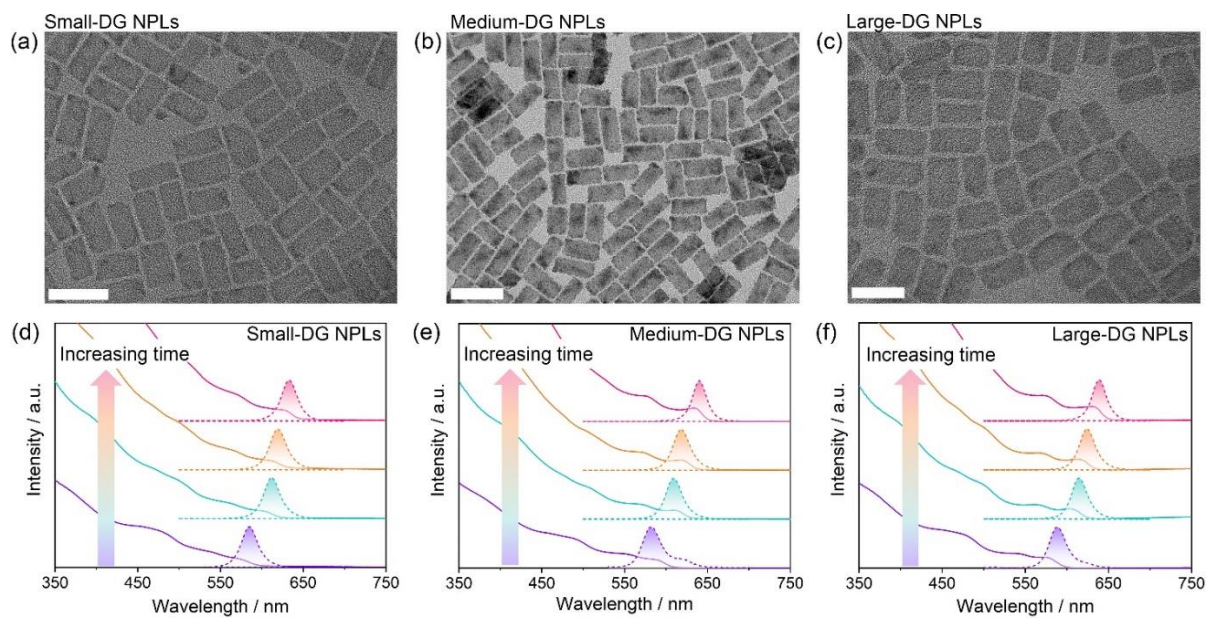

**Figure S2.** TEM images of the (a) small-, (b) medium-, and (c) large-DG NPLs (scale bar: 50 nm). The evolutions of the absorption and emission spectra of the (d) small-, (e) medium-, and (f) large-DG NPLs, monitored by taking small aliquots from the reaction solutions during the shell growth.

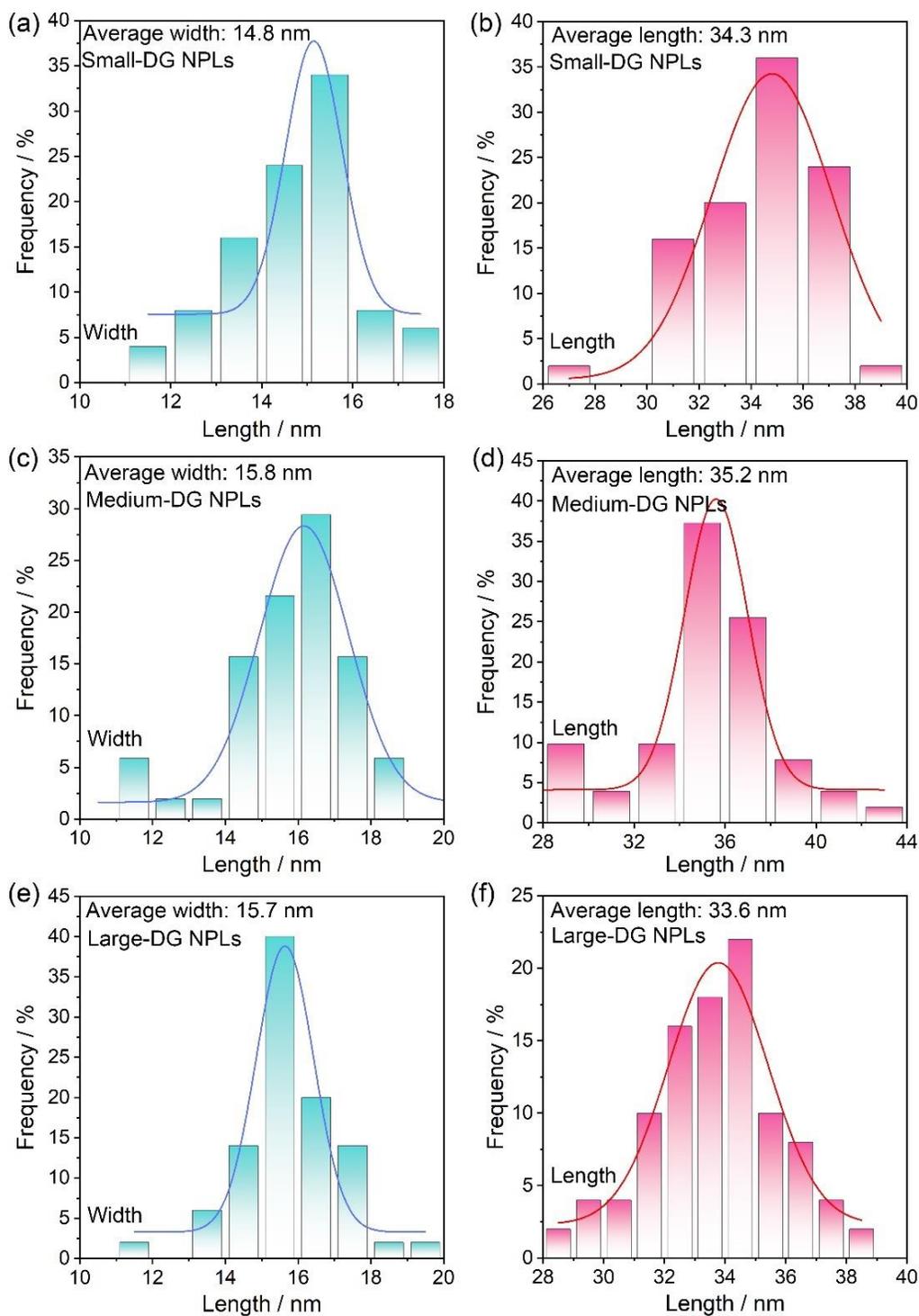

**Figure S3.** Size distributions of the width and length dimensions of (a,b) small-DG NPLs, (c,d) medium-DG NPLs, and (e,f) large-DG NPLs, respectively. The above results indicate that DG NPLs containing CdSe recombination centers of varying sizes exhibit comparable overall length and width dimensions.

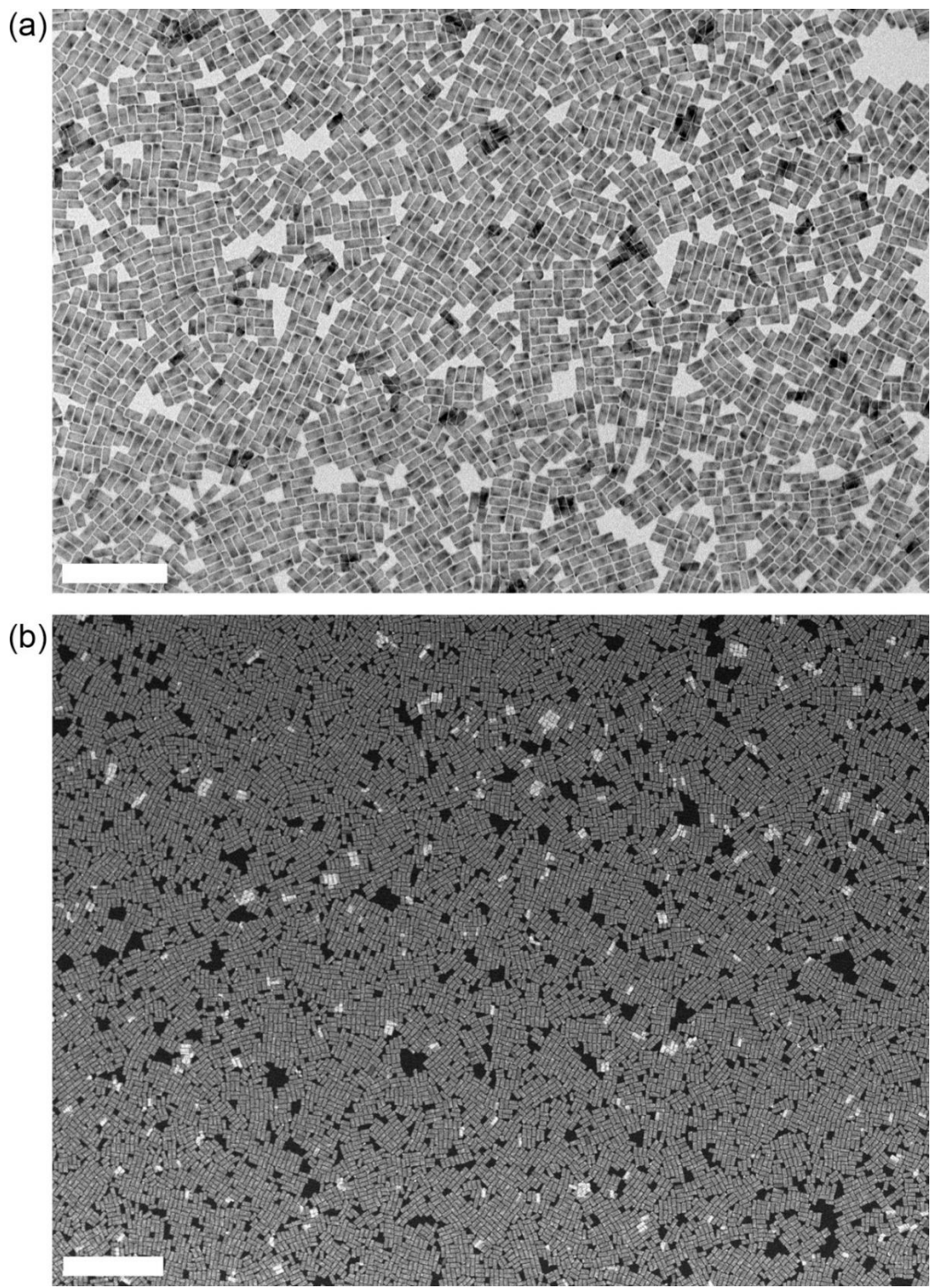

**Figure S4.** Representative large-area (a) TEM (scale bar: 200 nm) and (b) HAADF-STEM (scale bar: 400 nm) images of medium-DG NPLs, demonstrating the resulting DG NPLs after shell growth exhibit excellent uniformity in both lateral size and thickness.

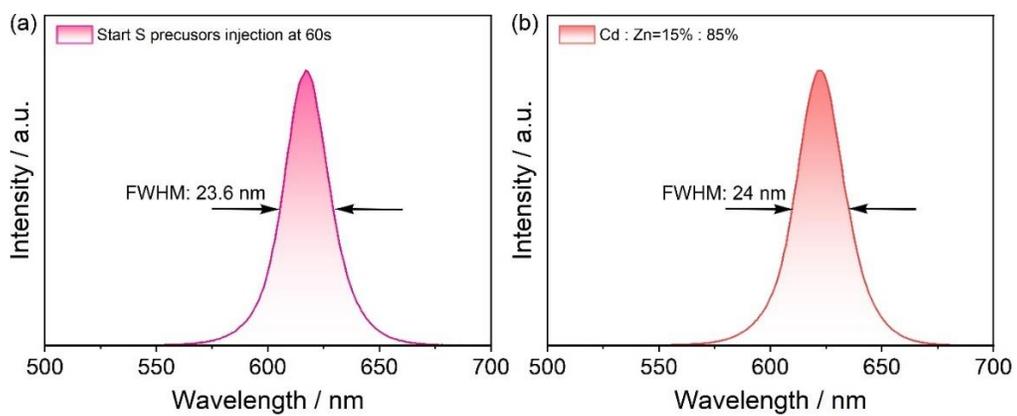

**Figure S5.** PL spectra of the resulting DG NPLs (a) using gradient cores with an earlier injection time of S precursors at 60 s and (b) by increasing the molar ratio of cations from Cd : Zn=25% : 75% to Cd : Zn=15% : 85% during the shell growth.

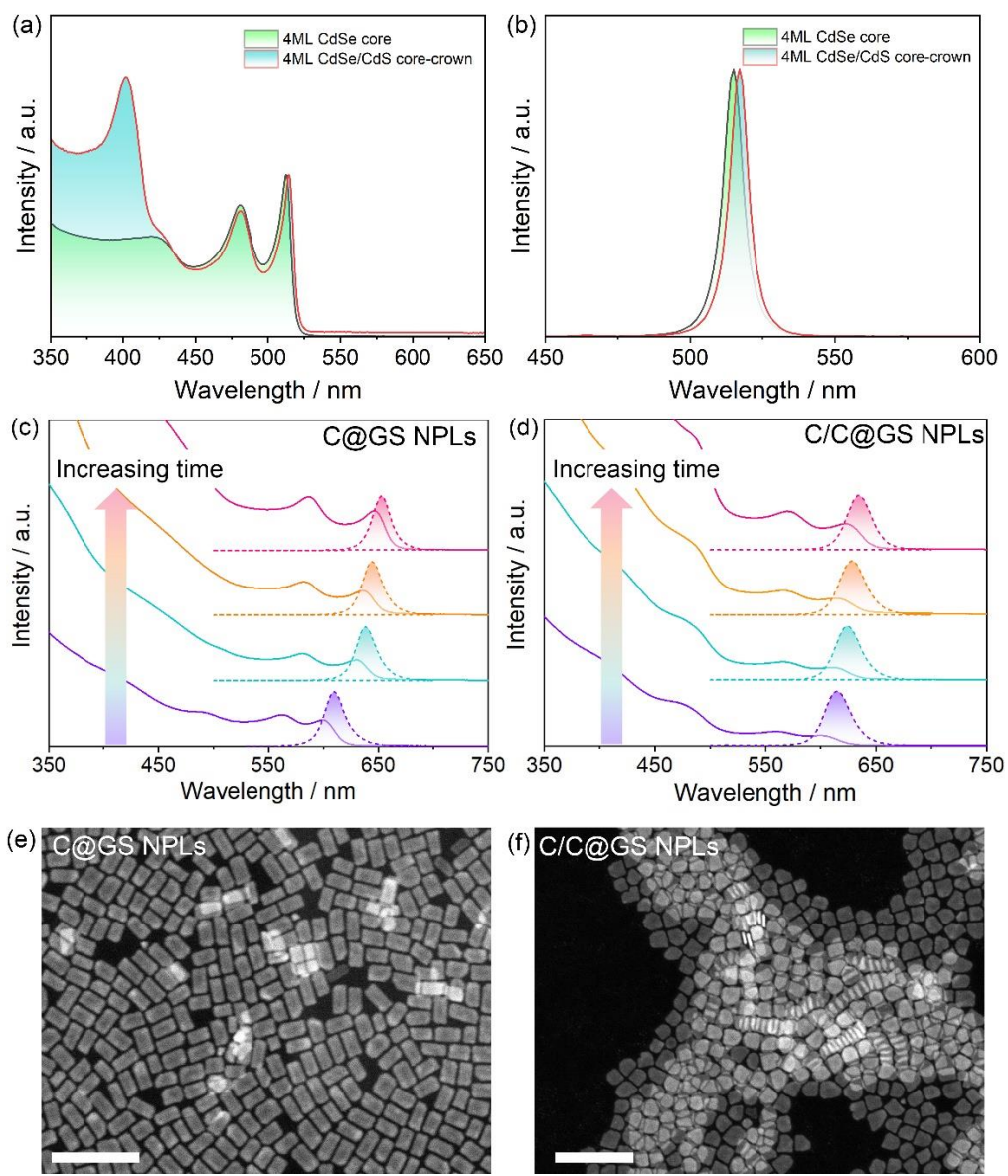

**Figure S6.** (a) Absorption and (b) PL spectra of the 4ML CdSe cores and 4ML CdSe/CdS core/crown NPLs using the conventional growth methods. The evolutions of the absorption and emission spectra of the (c) C@GS NPLs and (d) C/C@GS NPLs, monitored by taking small aliquots from the reaction solution during the shell growth. HAADF-STEM images of the resulting (e) C@GS NPLs and (f) C/C@GS NPLs after shell growth (scale bar: 200 nm). These results are highly consistent with the previous reports of core-shell structured NPLs. [3, 7, 8]

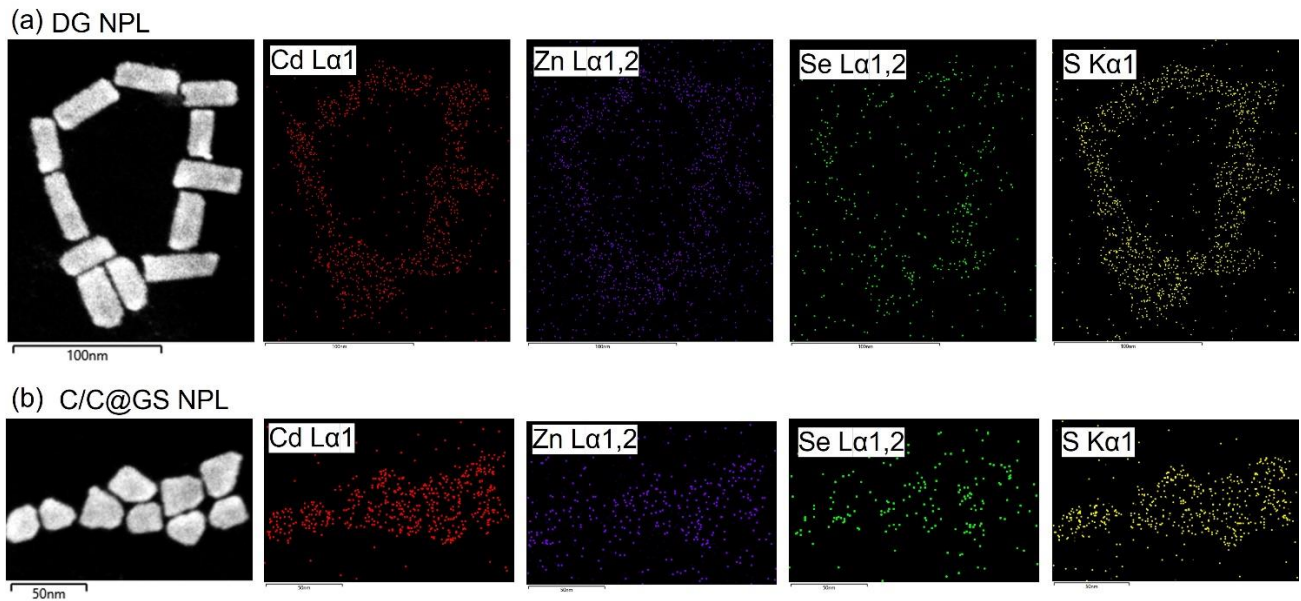

**Figure S7.** EDS mapping results of the (a) DG NPLs and (b) C/C@GS NPLs.

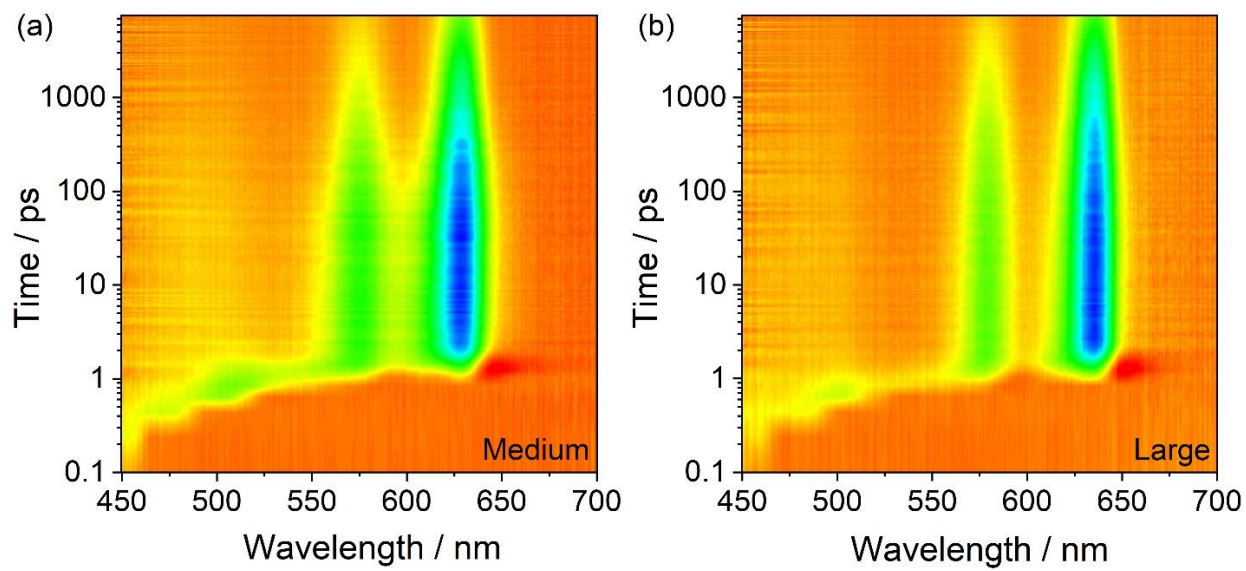

**Figure S8.** Transient absorption (TA) spectroscopies of (a) medium-DG NPLs and (b) large-DG NPLs. Similar to the results observed for small-DG NPLs, medium-, and large-DG NPLs also demonstrate highly efficient and ultrafast carrier transfer moving towards CdSe recombination center, with almost no trapped excitons in the $CdSe_xS_{1-x}$ region (450-550 nm).

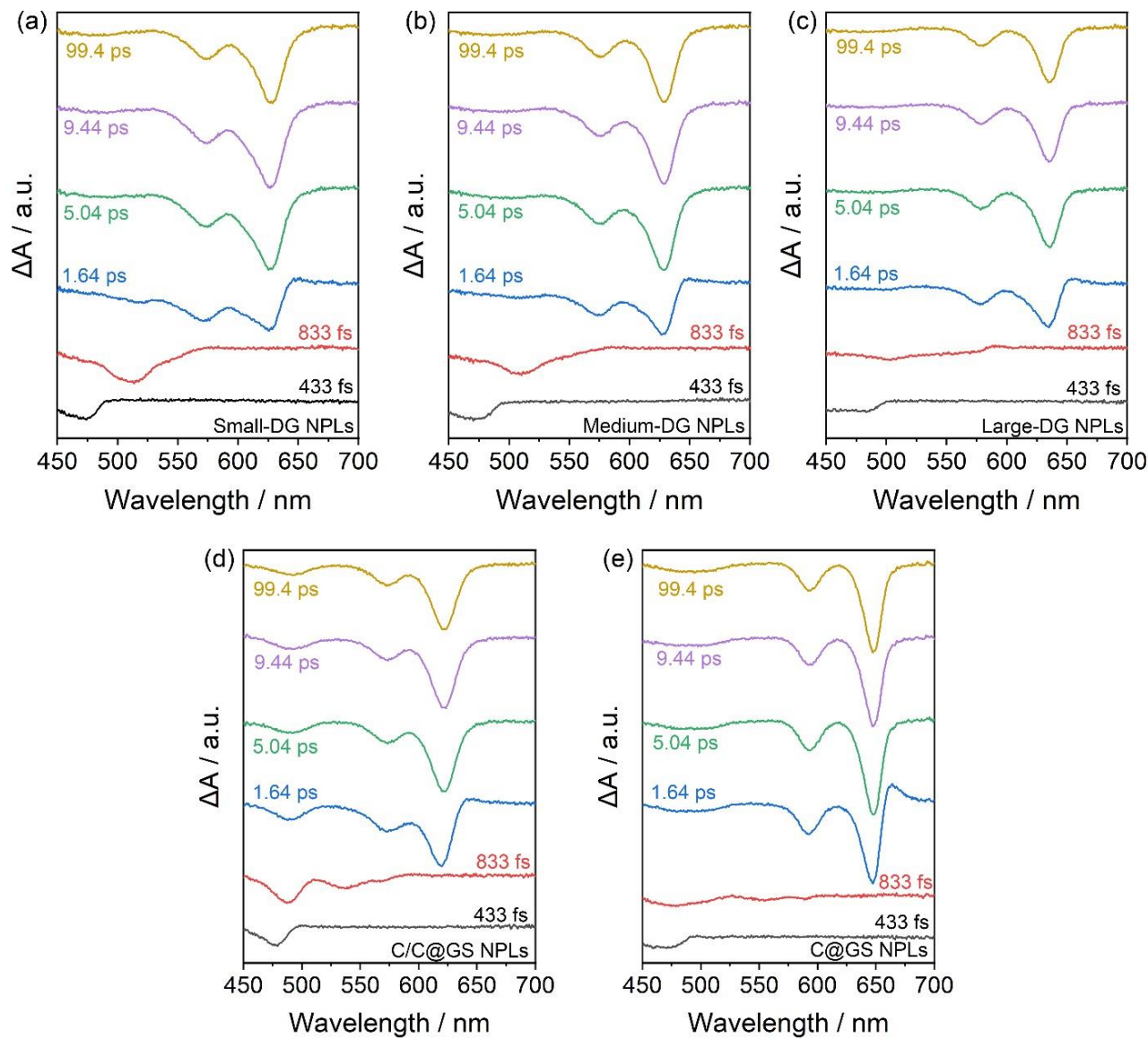

**Figure S9.** TA spectra of the (a) small-DG NPLs, (b) medium-DG NPLs, (c) large-DG NPLs, (d) C/C@GS NPLs, and (e) C@GS NPLs, reveal the evolution of ΔA at different decay intervals. As the relative size of CdSe become smaller, the bleach singles at 450-550 nm, corresponding to the excitons generated in the $CdSe_xS_{1-x}$ region, become stronger, reaching their maximum within the initial 833 fs. Afterward, the bleach singles of the light and heavy holes of CdSe become more intensive, accompanied by a fading of the 450-550 nm bleach single. The above results indicates that, at a consistent overall size of the DG NPLs, the smaller the size of the CdSe recombination center, the higher the exciton concentration ultimately achieved at the center. For conventional C/C@GS NPLs, controlling the concentration of the excitons in the recombination center is difficult, as the separation and purification of sub-10 nm 4ML CdSe cores are difficult to achieve synthetically. Furthermore, it can be observed that even after 99.4 ps, there is still a relatively strong bleach signal within the 450-550 nm range. This suggests that a large portion of excitons is trapped at the interface between CdS and CdSe, preventing the quantum efficiency from reaching near-unity. For C@GS NPLs, exciton generation and recombination are highly spatial overlapped within CdSe, leading to a significant self-absorption issue.

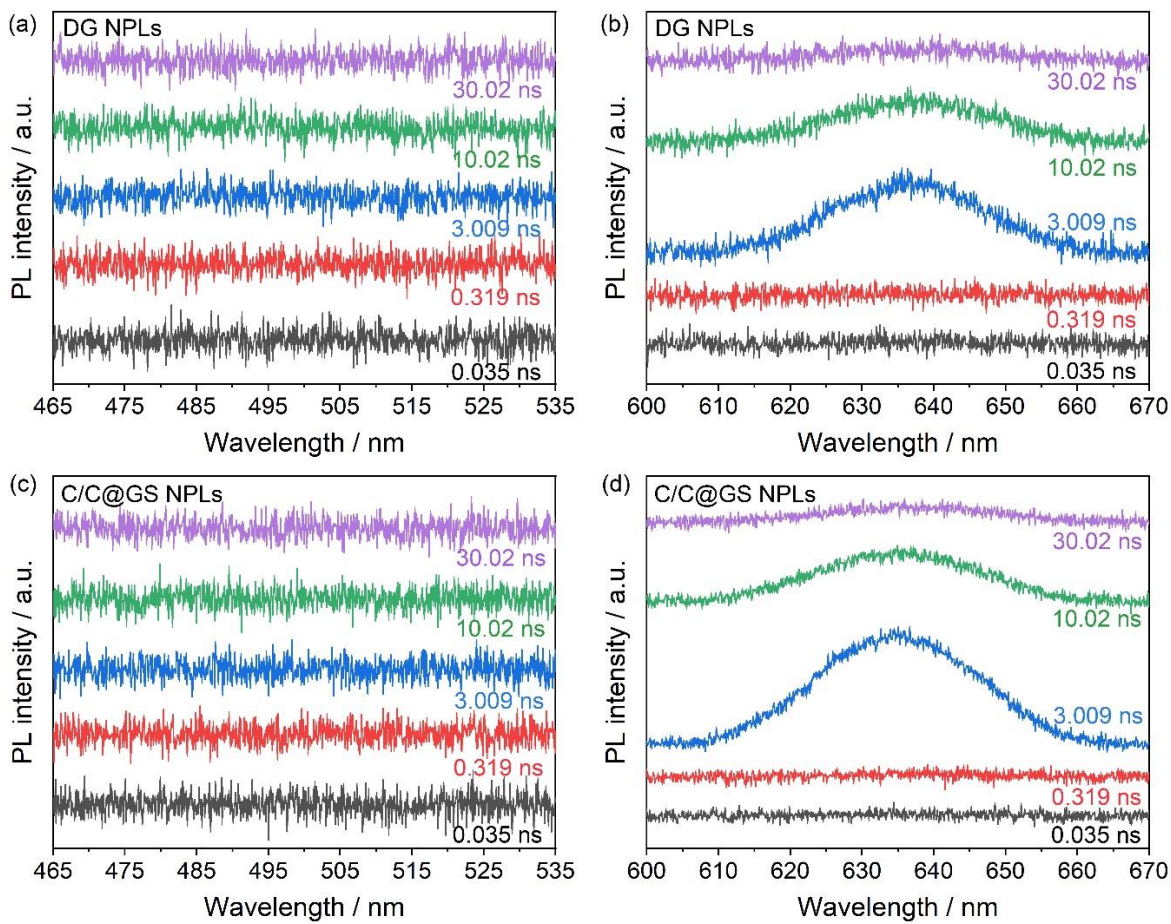

**Figure S10.** Time-resolved PL spectra of (a,b) DG NPLs and (c,d) C/C@GS NPLs at different decay intervals. These results demonstrate that all the emitted photons originate from exciton recombination within the CdSe regions. Moreover, it also confirmed that the slow decay observed in the 450-550 nm range in the TA spectra of C/C@GS NPLs can be attributed to non-radiative trap states.

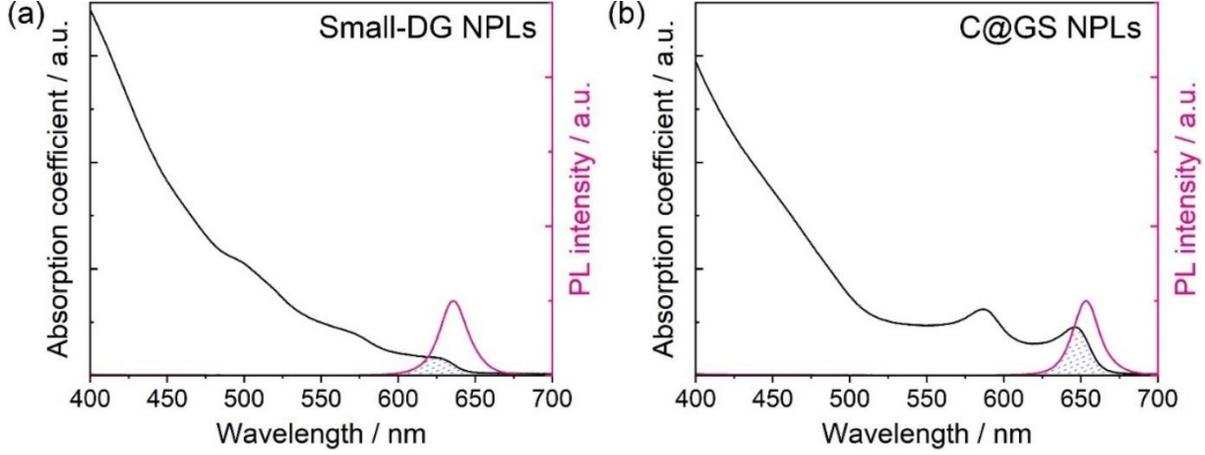

**Figure S11.** Absorption coefficient and PL spectra of (a) small-DG NPLs and (b) C@GS NPLs.

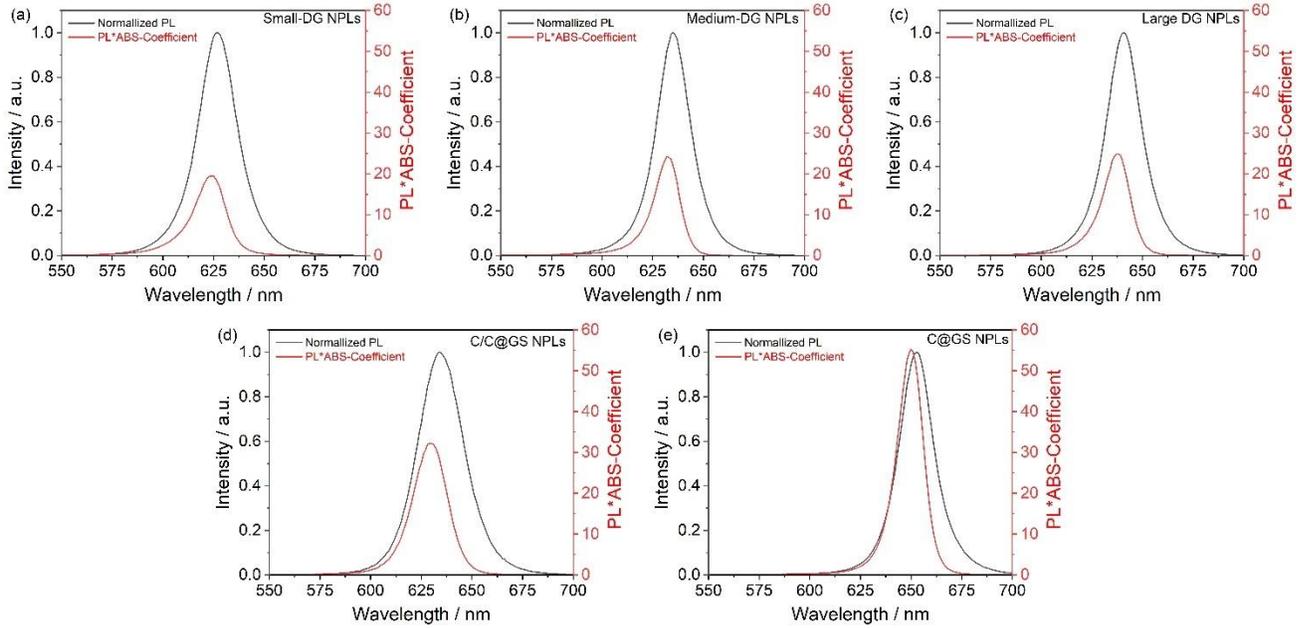

**Figure S12.** $S_{PL}(\lambda)\alpha(\lambda)$ spectra (red curves) and normalized PL spectra (black curves) of (a) small-DG NPLs, (b) medium-DG NPLs, (c) large-DG NPLs, (d) C/C@GS NPLs, and (e) C@GS NPLs.

The relative self-absorption coefficients of different samples are calculated and compared with each other according to the literature.[9] Specifically, the overlap between absorption coefficient spectra and PL spectra, as depicted in **Figure S11** using Small-DG NPLs and C@GS NPLs as examples, can be quantified as self-absorption coefficient $\alpha_s$, according to **Equation 1**. Here $S_{PL}(\lambda)$ and $\alpha(\lambda)$ refer to PL spectra and absorption coefficient spectra, respectively.

$$\alpha_S = \frac{\int S_{PL}(\lambda)\alpha(\lambda)\,d\lambda}{\int S_{PL}(\lambda)\,d\lambda} \qquad (\textbf{Equation 1})$$

The calculated $S_{PL}(\lambda)\alpha(\lambda)$ in 550-700 nm for all samples are shown in the red curves in **Figure S12**, in comparison to their corresponding normalized PL spectra. It can be clearly seen that the higher overlap between absorption coefficient and PL spectra, the greater the weight of $S_{PL}(\lambda)\alpha(\lambda)$ in the PL spectra. Finally, $\alpha_s$ of different samples are calculated and compared with C@GS NPLs according to Equation 1 based on the data in **Figure S12**.

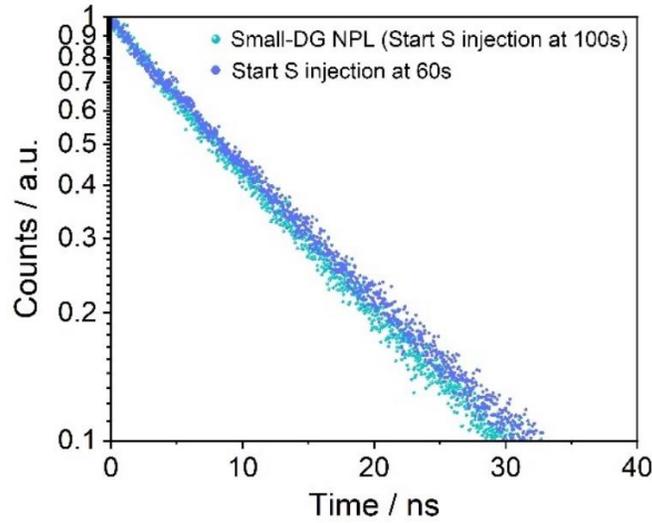

**Figure S13.** PL decay curves of small-DG NPL (start S precursors injection at 100 s, cyan dots) and DG NPLs with even smaller CdSe recombination center (start S precursors injection at 60 s, bule dots).

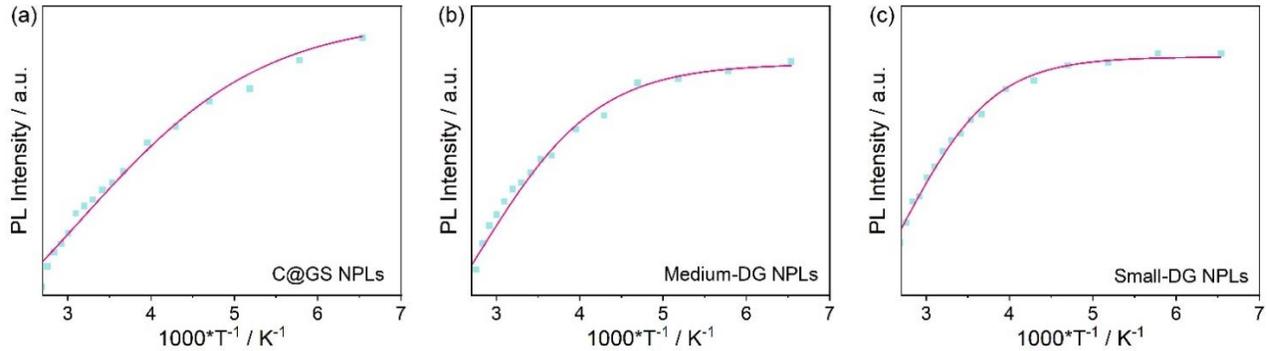

**Figure S14.** Integrated PL intensity as a function of temperature for (a) C@GS NPLs, (b) medium-DG NPLs, and (c) small-DG NPLs. The experimental data (cyan solid dots) are well-fitted by using **Equation 2** (magenta curves), as indicated by the literature.[10] The obtained $E_b$ for C@GS NPLs, medium-DG NPLs, and small-DG NPLs is 78.2, 121.7, and 158.9 meV, respectively.

$$I(T) = \frac{I_0}{1+Ae^{(-E_b/k_BT)}} \qquad \textbf{(Equation 2)}$$

Here $I_0$ refers to the PL intensity at 0 K, $k_B$ is the Boltzmann constant and $E_b$ is the exciton binding energy.

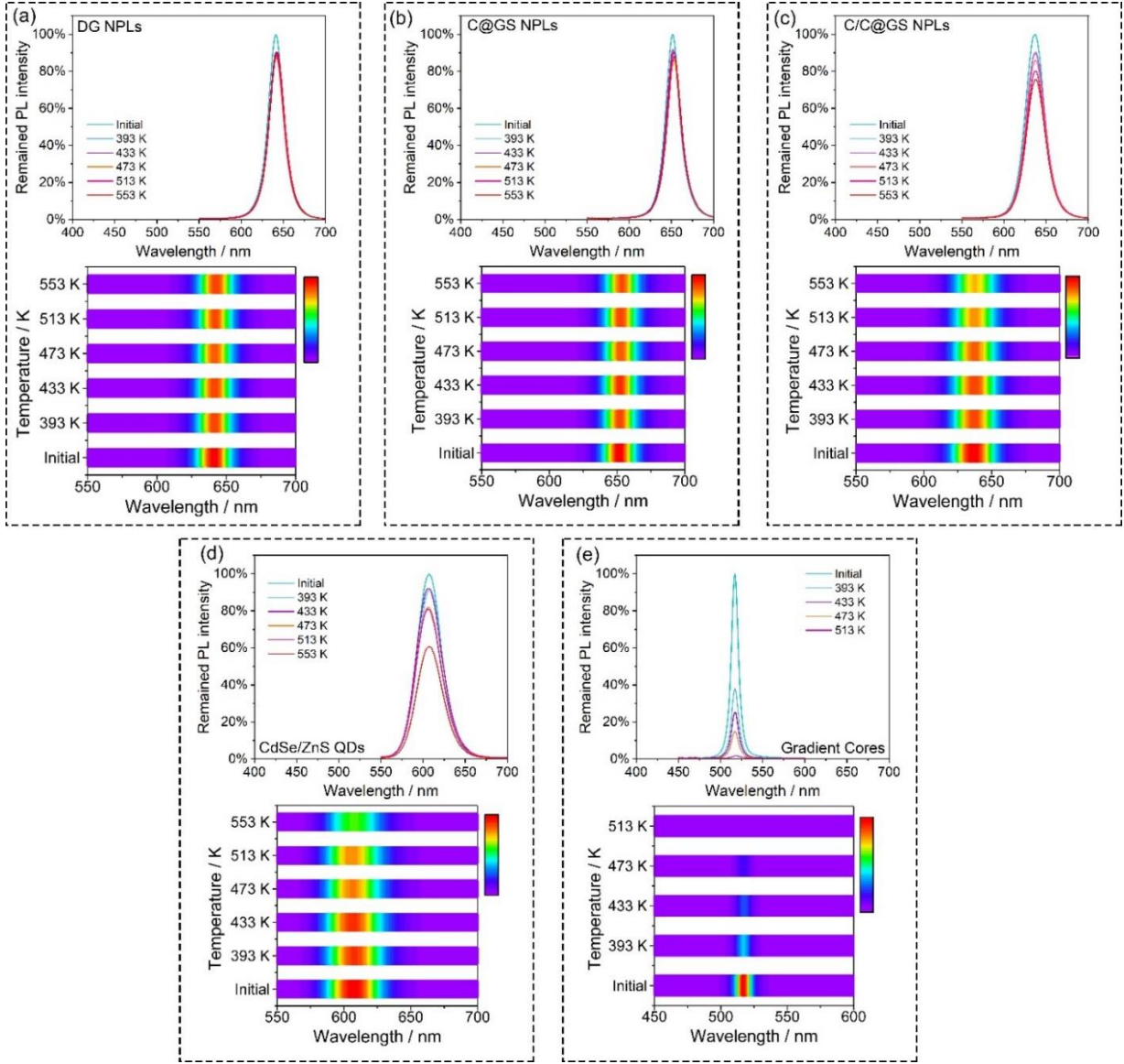

**Figure S15.** Thermal stability measurements of (a) DG NPLs, (b) C@GS NPLs, (c) C/C@GS NPLs, (d) CdSe@ZnS QDs, and (e) gradient cores by baking the samples at different elevated temperatures in the glove box for 10 min. The PL spectra of the samples were measured after the samples were cooled to room temperature.

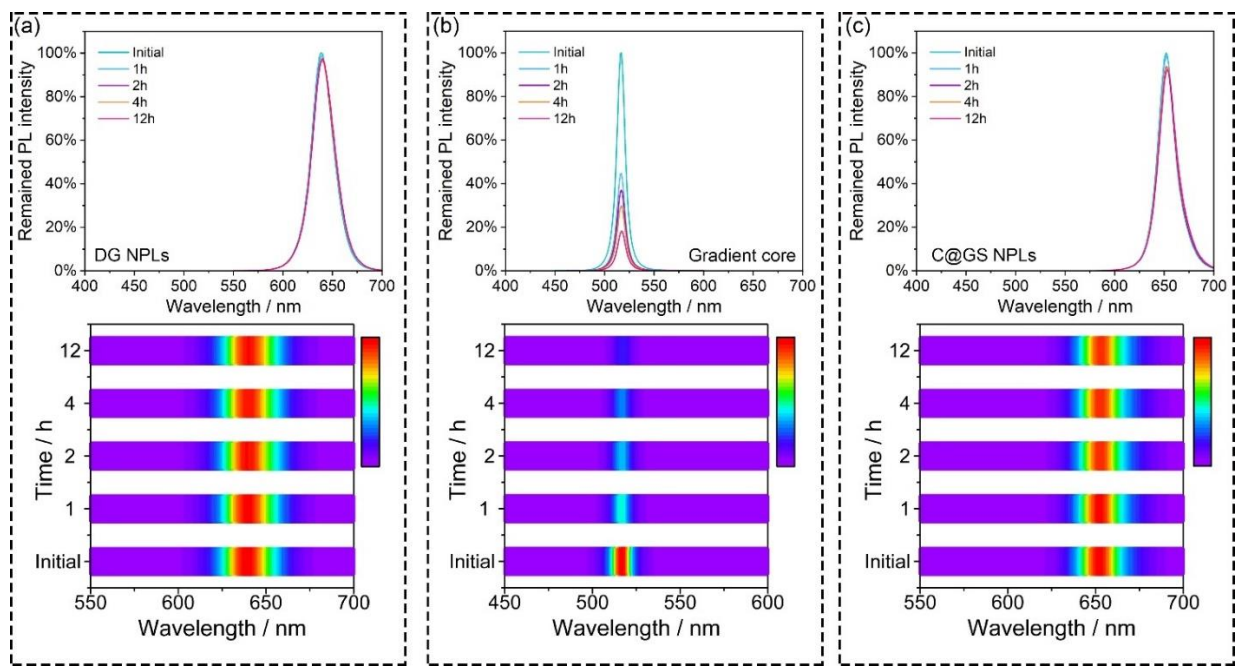

**Figure S16.** Long-term thermal stability measurements of (a) DG NPLs, (b) gradient core NPLs, and (c) C@GS NPLs by baking the samples in the air at 353 K. The PL spectra of the samples were measured at different time intervals during the baking process after the samples were cooled to room temperature.

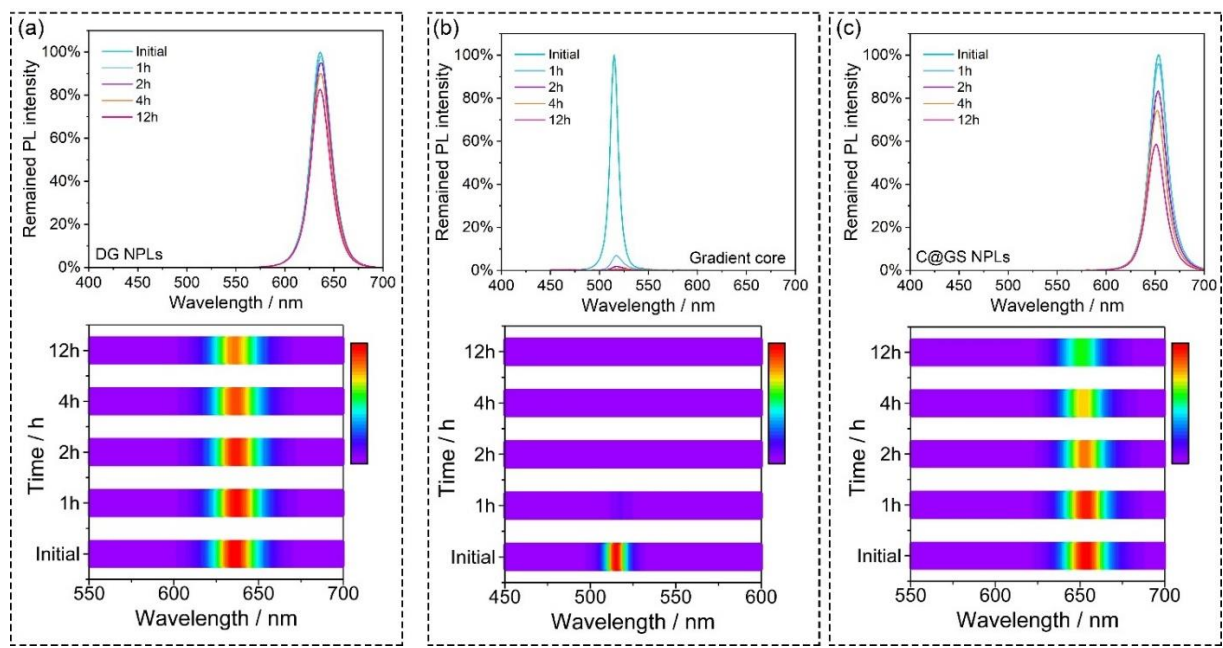

**Figure S17.** Long-term thermal stability measurements of (a) DG NPLs, (b) gradient core NPLs, and (c) C@GS NPLs by baking the samples in the air at 393 K. The PL spectra of the samples

were measured at different time intervals during the baking process after the samples were cooled to room temperature.

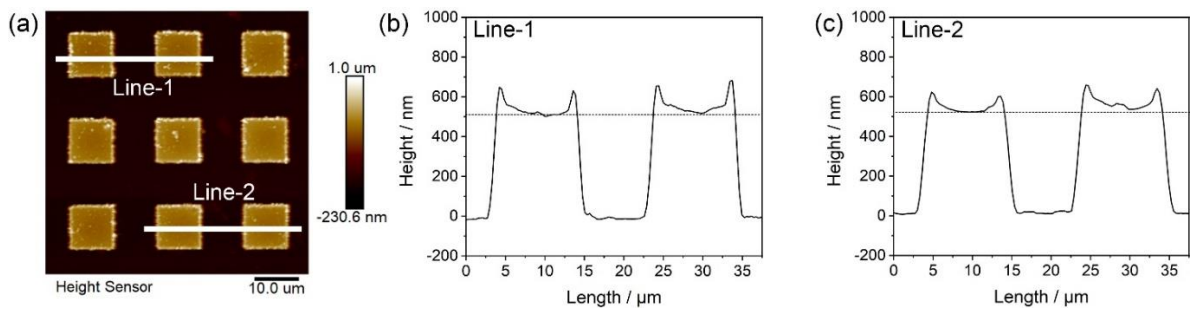

**Figure S18.** (a) AFM image of the DG NPLs pixels. The surface profiles of the DG NPLs pattern for (b) line-1 and (c) line-2 in the AFM image. The results show the large thickness of ~500 nm for the resulting NPLs pattern.

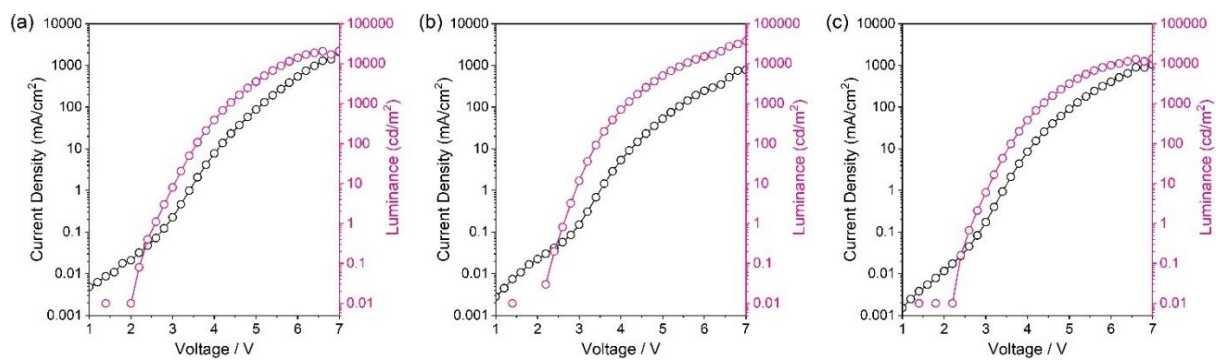

**Figure S19.** Current density and luminance of (a) small-DG NPLs, (b) medium-DG NPLs, and (c) C@GS NPLs at different applied voltages.

**Reference**

1. Altintas Y, *et al.* Highly Stable, Near-Unity Efficiency Atomically Flat Semiconductor Nanocrystals of CdSe/ZnS Hetero-Nanoplatelets Enabled by ZnS-Shell Hot-Injection Growth. *Small* **15**, 1804854 (2019).

2. Bertrand GH, Polovitsyn A, Christodoulou S, Khan AH, Moreels I. Shape control of zincblende CdSe nanoplatelets. *ChemComm* **52**, 11975-11978 (2016).

3.  Taghipour N, *et al.* Sub-single exciton optical gain threshold in colloidal semiconductor quantum wells with gradient alloy shelling. *Nat. Commun.* **11**, 3305 (2020).

4.  Lim J, Jun S, Jang E, Baik H, Kim H, Cho J. Preparation of highly luminescent nanocrystals and their application to light-emitting diodes. *Adv. Mater.* **19**, 1927-1932 (2007).

5.  Hu S, *et al.* High-performance deep red colloidal quantum well light-emitting diodes enabled by the understanding of charge dynamics. *ACS Nano* **16**, 10840-10851 (2022).

6.  Gheshlaghi N, *et al.* Self-resonant microlasers of colloidal quantum wells constructed by direct deep patterning. *Nano Lett.* **21**, 4598-4605 (2021).

7.  Samadi Khoshkhoo M, Prudnikau A, Chashmejahanbin MR, Helbig R, Lesnyak V, Cuniberti G. Multicolor patterning of 2D semiconductor nanoplatelets. *ACS Nano* **15**, 17623-17634 (2021).

8.  Shabani F, *et al.* Deep-Red-Emitting Colloidal Quantum Well Light-Emitting Diodes Enabled through a Complex Design of Core/Crown/Double Shell Heterostructure. *Small* **18**, 2106115 (2022).

9.  Klimov VI, Baker TA, Lim J, Velizhanin KA, McDaniel H. Quality factor of luminescent solar concentrators and practical concentration limits attainable with semiconductor quantum dots. *ACS Photonics* **3**, 1138-1148 (2016).

10. Li F, *et al.* Enhancing exciton binding energy and photoluminescence of formamidinium lead bromide by reducing its dimensions to 2D nanoplates for producing efficient light emitting diodes. *Nanoscale* **10**, 20611-20617 (2018).